\documentclass[sigconf]{acmart}

\usepackage{tikz}
\usepackage{amsmath}

\usepackage{stfloats}
\usepackage{filecontents}
\usepackage{verbatim}
\usepackage{enumerate}

\usepackage{bbding}
\usepackage{pifont}
\usepackage{wasysym}
 
\usepackage{amssymb}

\usepackage{algorithm} 

\usepackage{algorithmic} 

\usepackage{multirow} 

\usepackage{amsmath} 

\usepackage{xcolor}

\usepackage{epsfig} 
\usepackage{setspace}
\usepackage{balance}
\usepackage{verbatim}
\usepackage{amsfonts}
\usepackage{epsfig}  
\usepackage{amssymb}
\usepackage[normalem]{ulem}
\usepackage{amscd}
\usepackage{amsmath} 
\usepackage{microtype}
\usepackage{array}
\usepackage{chngpage}
\usepackage{graphicx}
\usepackage{epsfig}
\usepackage{multirow}
\usepackage{caption}
\usepackage{listings}
\usepackage{subfigure}
\usepackage{adjustbox}
\usepackage{ulem}
\usepackage{lipsum}
\usepackage{float}
\usepackage{pifont}
\usepackage{bm}
\usepackage{color,xcolor}
\usepackage{booktabs} 
\usepackage{verbatim}
\usepackage{amssymb}
\usepackage{pifont}
\usepackage{array,multirow}
\usepackage{makecell}
\usepackage{tikz}
\usepackage{CJK}
\usepackage{appendix}  
\usepackage{float}
\usepackage{threeparttable}
\usepackage{colortbl}
\usepackage{titlesec}
\usepackage{appendix}
\usepackage{cleveref}
\usepackage[utf8]{inputenc}

\crefname{section}{§}{§§}
\Crefname{section}{§}{§§}
\usepackage{fancyhdr}

\newcommand{\ignore}[1]{}

\newcommand\kai[1]{{\textcolor{blue}{Kai: #1}}}
\newcommand\shengzhi[1]{{\textcolor{red}{Shengzhi: #1}}}

\AtBeginDocument{%
  \providecommand\BibTeX{{%
    \normalfont B\kern-0.5em{\scshape i\kern-0.25em b}\kern-0.8em\TeX}}}

\setcopyright{acmcopyright}
\copyrightyear{2018}
\acmYear{2018}
\acmDOI{10.1145/1122445.1122456}

\begin{document}
\cfoot{\thepage}

\title{HufuNet: Embedding the Left Piece as Watermark and Keeping the Right Piece for Ownership Verification in Deep Neural Networks}

\author{Peizhuo Lv$^{1,2}$, Pan Li$^{1,2}$, Shengzhi Zhang$^{3}$, Kai Chen$^{1,2}$, Ruigang Liang$^{1,2}$, Yue Zhao$^{1,2}$, Yingjiu Li$^{4}$}

\affiliation{\normalsize{$^{1}$SKLOIS, Institute of Information Engineering, Chinese Academy of Sciences, China}}
\affiliation{\normalsize{$^{2}$School of Cyber Security, University of Chinese Academy of Sciences, China}}
\affiliation{\normalsize{$^{3}$Department of Computer Science, Metropolitan College, Boston University, USA}}
\affiliation{\normalsize{$^{4}$University of Oregon, USA}}

\email{{lvpeizhuo, lipan}@iie.ac.cn, shengzhi@bu.edu, {chenkai, liangruigang, zhaoyue}@iie.ac.cn, yingjiul@uoregon.edu}

\thispagestyle{empty}
\renewcommand{\headrulewidth}{0pt}
\renewcommand{\footrulewidth}{0pt}

\settopmatter{printacmref=false} 
\renewcommand\footnotetextcopyrightpermission[1]{} 
\pagestyle{plain}



\begin{abstract}

Due to the wide use of highly-valuable and large-scale deep neural networks (DNNs), it becomes crucial to protect the intellectual property of DNNs so that the ownership of disputed or stolen DNNs can be verified. Most existing solutions embed backdoors in DNN model training such that DNN ownership can be verified by triggering distinguishable model behaviors with a set of secret inputs. However, such solutions are vulnerable to model fine-tuning and pruning. They also suffer from fraudulent ownership claim as attackers can discover adversarial samples and use them as secret inputs to trigger distinguishable behaviors from stolen models. To address these problems, we propose a novel DNN watermarking solution, named $HufuNet$\footnote{Hu-fu, also known as tiger tally, was used by an emperor to command the army in ancient China. It was divided into two pieces, with the emperor keeping the right piece and the local commander keeping the left. When a new general started to command the army on behalf of the emperor, he/she was required to present the right piece that matched the left one from the local commander~\cite{hufutally}.}, for protecting the ownership of DNN models. We evaluate HufuNet rigorously on four benchmark datasets with five popular DNN models, including convolutional neural network (CNN) and recurrent neural network (RNN). The experiments demonstrate HufuNet is highly robust against model fine-tuning/pruning, kernels cutoff/supplement, functionality-equivalent attack, and fraudulent ownership claims, thus highly promising to protect large-scale DNN models in the real-world. 

\end{abstract}

\maketitle
\vspace{-4pt}
\section{Introduction}
\label{sec:Introduction}

The rapid development of artificial intelligence and machine learning technologies in recent years has driven the broad adoption of deep neural networks (DNNs) in numerous applications such as computer vision~\cite{he2016deep,he2016identity,simonyan2014very}, natural language processing~\cite{wu2016google,collobert2011natural,xiong2016achieving}, and speech recognition~\cite{amodei2016deep,ko2015audio}. DNNs are valuable intellectual property (IP) of their owners due to the tremendous amount of expertise, effort, and computing power involved in the design, training, validating, and commercializing them. Such IP should be protected according to the IP protection law. For example, the European Patent Office has recently amended its ``Guidelines for Examination'' on how patents related to artificial intelligence and machine learning technologies should be assessed~\cite{lankinen2020patentability}. 

Attackers with full access to the structures and parameters of DNN models may steal and possibly modify them and fraudulently claim ownership. Such IP stealing may happen in various real-world settings such as: (i) Owners outsource the training of their DNN models to third parties, (ii) Owners upload their DNN models and services to cloud service providers, and (iii) Owners license their DNN models and services to third parties. Attackers may achieve their goals via malware infection~\cite{watson2015malware,jamil2011security}, insider threats~\cite{claycomb2012insider,silowash2012common}, industrial espionage, and many other ways, leading to serious IP infringement and significant economic loss to owners.




Watermarking, a process of hiding digital information in carrier data, has been used for media rights management~\cite{chen2001quantization,hartung1999multimedia} and software ownership protection~\cite{collberg1999software,venkatesan2001graph}. It has also been developed for the IP protection of DNNs in recent years. Most DNN watermarking techniques embed backdoors~\cite{adi2018turning,zhang2018protecting,namba2019robust,guo2018watermarking,rouhani2018deepsigns,li2019prove} into DNN models such that watermarked DNNs behave in a designed, distinguishable manner given specific inputs held by model owners as secret triggers. However, recent works \cite{chen2019leveraging,liu2018fine} show that fine-tuning~\cite{yosinski2014transferable,pittaras2017comparison} and pruning~\cite{han2015learning} can eliminate the embedded backdoors (i.e., watermarks) without significant performance loss or dramatic change of the protected models. Most existing DNN watermarking techniques are also vulnerable to fraudulent ownership claims---attackers may discover adversarial samples and use them as ``secret triggers''. It is known that adversarial samples are inherent to DNN models and can cause the models to demonstrate distinguishable behaviors ~\cite{szegedy2013intriguing,biggio2013evasion}. Attackers can claim ownership over the stolen models by querying them with adversarial samples.   


\vspace {3pt}\noindent\textbf{Challenges in watermarking DNNs}. First, watermarks could be eliminated/undermined by attackers via fine-tuning, pruning, or kernels cutoff/supplement, by which attackers change model parameters or the number of kernels but maintain the performance of the watermarked models. The change of model parameters may effectively remove the embedded watermarks as shown in previous studies on DNN watermarking techniques~\cite{adi2018turning,uchida2017embedding,zhang2018protecting,guo2018watermarking,rouhani2018deepsigns}. Second, it is challenging to design a robust watermark without significantly sacrificing the performance of the original model. Strong robustness tends to embed a large amount of watermark information into DNN models, downgrading the watermarked models' performance to an unacceptable level. In specific applications such as autonomous driving, even 1\% of performance loss is intolerable. It is by no means trivial to find a way of embedding enough watermark information for strong robustness while keeping high fidelity of DNN models. Third, it is also challenging to prevent attackers from forging new watermarks and claim ownership fraudulently over stolen DNN models. In backdoor-based watermarking, owners of DNN models keep secret triggers as ownership proofs. However, attackers can discover adversarial samples as their triggers and confuse DNN ownership judgment effectively. Such attacks are difficult to address under the assumption that attackers understand the underlying watermarking mechanisms and explore all possible ways in forging valid watermarks to claim ownership over stolen models.

\vspace {3pt}\noindent\textbf{HufuNet Approach}. To address these challenges, we propose $HufuNet$, a novel white-box watermarking scheme for DNN models. 
HufuNet approach can be divided into three phases: watermark generation, watermark embedding, and ownership verification. 

In the watermark generation phase, a neural network with a small number of parameters, called HufuNet, is trained to yield a high accuracy when it is tested. The set of training samples and testing samples are public references that must be used later in the ownership verification phase. 
After HufuNet is trained and tested, it is split into two pieces, with the left piece as a watermark to be embedded into a DNN model for ownership protection, and the right kept by the model owner as a secret for ownership verification.
 
In the watermark embedding phase, each convolution kernel of the left piece of HufuNet is embedded into the target DNN model's architecture that consists of a large number of parameters. The location to embed each convolution kernel is computed and selected to increase watermark forgery attack difficulty. 
During the watermarked DNN model training, the parameters incorporated from HufuNet are frozen while the rest of the parameters in the model are updated. This watermark embedding process ensures that the watermarked DNN embeds the left piece of HufuNet, yet it is fully trained on its own dataset to preserve the performance on its main task. 

In the ownership verification phase, the model's owner extracts the embedded watermark from a suspect DNN model at computed locations and recovers the left piece of HufuNet accordingly. Next, the owner brings out the right piece of HufuNet, merges it with the recovered left piece to form a whole HufuNet. The whole HufuNet is tested on the public set of testing samples (previously used in watermark generation). If and only if the accuracy change of the whole HufuNet (compared to the original HufuNet) is no more than a predetermined threshold, the ownership can be claimed over the suspect model. 

HufuNet is evaluated on five popular DNN models (including both CNN and RNN models) and four benchmark datasets. Experimental results show that HufuNet achieves state-of-the-art performance in the aspects as below. 
In particular, after the watermark is embedded, the accuracy of these DNN models does not decline, with a 0.23\% increase on average, achieving excellent fidelity compared to their non-watermarked versions under the same training settings.
The fine-tuning attack decreases the accuracy of the five watermarked DNNs by 1.20\% on average, but the accuracy of ownership verification decreases by only 0.36\%. When the watermarked DNN models' parameters are pruned from 10\% to 90\%, the accuracy of the watermarked DNN models may drop significantly (i.e., 61.10\% on average) to an unacceptable level. However, the accuracy of ownership verification for these pruned models drops insignificantly (i.e., 29.43\% on average). Regarding kernels cutoff attack, when 6.75\%, 2.30\%, and 4.77\% kernels are eliminated from VGG11, Resnet18, GoogLeNet respectively, we can still retrieve the remaining kernels correctly to rebuild HufuNet with watermark detection accuracy at 100\%. 
Moreover, our watermark is shown to be capable of dealing with functionality-equivalent (via structure adjustment, parameter adjustment or channel expansion) attacks without any performance downgrade. We also evaluate HufuNet against synthetic attacks, integrating fine-tuning, pruning, and functionality-equivalent attack together, and the accuracy of the retrieved HufuNet is still always above the threshold to verify the ownership as long as the adversaries would preserve the original functionality of the stolen model. The experimental results also demonstrate that statistical approaches (examining the distribution of parameter values and their gradients) cannot help learn the embedded watermark's existence. Finally, our case studies show that it is infeasible for the attacker to retrieve our HufuNet or forge a valid HufuNet for fraudulent ownership claim.



\vspace{5pt}\noindent\textbf{Contributions}. Our main contributions are outlined below:

\vspace {2pt}\noindent$\bullet$\space\textit{New technique.} We propose a new DNN watermarking approach, which embeds a piece of the watermark into a DNN model for ownership protection and withholds the other piece for ownership verification. To the best of our knowledge, this is the first approach that splits a watermark into two pieces and combines them for ownership verification.
 
\vspace {2pt}\noindent$\bullet$\space\textit{Implementation and evaluation.} We implement our watermark HufuNet and rigorously evaluate it using both CNN and RNN models. Experimental results show that our new DNN watermarking approach can simultaneously achieve robustness, adaptiveness, stealthiness, security, integrity, and fidelity. We release our HufuNet framework to the community\footnote{https://github.com/HufuNet/HufuNet}, together with the training and testing dataset for HufuNet, hoping to contribute to the IP protection of neural network models.


\vspace{-4pt}
\section{Background}
\label{sec:Background}

\ignore{\subsection{Deep Neural Networks}
A neural network can be viewed as a mapping function $f: \mathbb{R}^M \rightarrow \mathbb{R}^N$ given a training dataset $D_{tr} = \{(x^1_{tr}, y^1_{tr}), \cdots, (x^n_{tr}, y^n_{tr})\}$, where $x^i_{tr} \in \mathbb{R}^M$, $y^i_{tr} \in \mathbb{R}^N$. If it is well trained, $f$ should output $y^i$ with a high probability for the input $x^i$. Function $f$ can be computed based on a weight vector $w$ organized in a network structure, where the weight values are determined in the training process using the training data $D_{tr}$. If the structure of a neural network consists of multiple layers, it is usually called a deep neural network (DNN). DNN is the foundation of many AI applications, which drives the rapid development of speech and image recognition, natural language processing, autonomous driving, medical diagnosis, etc. It is usually highly costly to train a DNN model due to the resources it consumes, including a large amount of high-quality training data, high-performance computing facilities, as well as domain experts for designing network structures and training parameters. Therefore, it is necessary to protect the IP of DNN models using techniques such as digital watermarking. }

\subsection{Digital Watermarking}
Digital watermarking is a content-based, information hiding technique for embedding/detecting digital information into/from carrier data (e.g., multimedia, documents, software, databases, etc.). It is desired that the embedded watermark should not impact the regular use of carrier data, and be difficult to be detected or removed. Recently, digital watermarking has been applied to the protection of DNN models, and can be classified into white-box approaches and black-box approaches. The white-box approaches assume that the owner of a DNN model can access the ``internals'' of a suspect model to verify the ownership of the model. For instance, Uchida et al. use a parameter regularizer to embed a watermark, which is a matrix of parameters, into the target model’s parameter space during its training process. When verifying ownership over a suspect model, the owner retrieves the matrix from the model, computes the retrieved matrix's product and a pre-generated key vector, and compares the result with a local secret kept by the owner~\cite{uchida2017embedding}.

In contrast, black-box watermarking approaches mostly rely on probing a suspect model for ownership verification, thus relaxing the requirement of the internal knowledge of the suspect model. Black-box watermarking approaches can further be categorized into two classes, blind~\cite{li2019prove,namba2019robust} approaches and non-blind approaches~\cite{adi2018turning,guo2018watermarking,zhang2018protecting,rouhani2018deepsigns}, depending on whether or not an embedded watermark can be perceived by the human visual system~\cite{li2019prove}. 

Most black-box watermarking schemes are implemented based on embedding backdoors~\cite{chen2017targeted,gu2017badnets} into the target DNN models. Embedding a backdoor into a neural network model is to deliberately train the model using a training dataset that is augmented with certain specific inputs (i.e., triggers) and their desired labels (typically wrong labels), which should be selected uniquely by the model's owner and thus demonstrate the model's distinguishable behavior. Once trained, the model should output the desired labels for the specific inputs. The mapping between the specific inputs and their desired labels is considered as a backdoor, and used as a watermark. The specific inputs are kept as a secret by the model's owner. The model's owner can claim his/her ownership over a suspect model if the backdoor/watermark is detected in the sense that the model outputs the desired labels given the specific inputs from the owner's secret. However, recent research indicates that most backdoor-based watermarks are vulnerable to be detected and destroyed~\cite{chen2018detecting,liu2018fine,wang2019neural}.

\subsection{Watermark Destructing Approaches}
\label{sub:destructing}
\noindent\textbf{Entire model retraining.} The most effective method to destruct an embedded watermark in a model is to retrain it completely. However, since such entire model retraining typically requires similar amount of expertise, effort, training data and computing power as training their own model from scratch, adversaries would rather train their own model than stealing to avoid any potential legal issues.  

\noindent\textbf{Fine-tuning.} Fine-tuning is typically used to optimize an already-trained DNN model to augment it with additional tasks or reinforce its performance. Fine-tuning involves an update to the parameter values in the model, thus possible destroying the embedded watermark. Generally, the scale of fine-tuning is flexible, depending on the training dataset and computing power available. Therefore, adversaries can fine-tune the stolen model according to their capacity to destruct the watermark.

\noindent\textbf{Pruning.} Model pruning approach is commonly used for model compression to produce smaller, more memory-efficient and power-efficient models with negligible loss inaccuracy. Typically, it prunes less-connected, unimportant neurons in the neural network by zeroizing them without changing the network structure. Model pruning can be used by those adversaries lack of training power to destruct the embedded watermark in their stolen model, since it does not involve any retraining.

\noindent\textbf{Functionality-equivalent attack.} Functionality equivalent attack can be launched via structure adjustment and parameter adjustment. The attackers do not need any training resources or training dataset to retrain the model. Instead, they can simply reorder the sequence of output channels on one convolution layer, and adjust the corresponding sequence of input channels on the next convolution layer to obtain a functionality-equivalent model (called structure adjustment in this paper), due to the isomorphism of many CNN structures. Furthermore, if the stolen DNN model is with ReLU as the activation function, the attackers can also scale the weights of all the neurons on one layer by a positive constant $c \textgreater 0$, and then scale the next layer by $1/c$ to get a functionality-equivalent model (called parameter adjustment in this paper) to destruct embedded watermarks. Functionality-equivalent attack can also be launched via channel expansion. In particular, adversaries randomly choose two adjacent layers of their stolen model with matrices $A$ and $B$. Then they increase the output dimension of $A$ (by increasing its output channels) to get $A'$, and $B$ to get $B'$, with the restriction of $(A' B') x = (A B) x$, where $x$ is the feature map propagated in neural networks. Finally, the adversaries can further apply structure adjustment or fine-tuning to undermine the embedded watermarks.


\noindent\textbf{Kernels cutoff and supplement.} The adversary can cut off (completely remove some kernels, rather than zeroizing them as pruning) or supplement some kernels in the stolen model to obtain a new model with similar performance but different number of kernels and structure. To ensure similar performance as the stolen model, the adversary should cut off or supplement kernels carefully: if some output channels (kernels) of one layer are cut off or supplemented, then the corresponding input channel of the next layer should also be cut off or supplemented. Besides, since kernels cutoff will reduce the performance of the model, the adversary may not want to cut off too many kernels from the stolen model. Kernels cutoff or supplement can also be leveraged by the adversary to destruct embedded watermarks. 



\vspace{-4pt}
\section{Problem Statement}
\label{sec:Motivation}

\subsection{A Motivating Example}
\begin{figure}[t]
\centering
\epsfig{figure=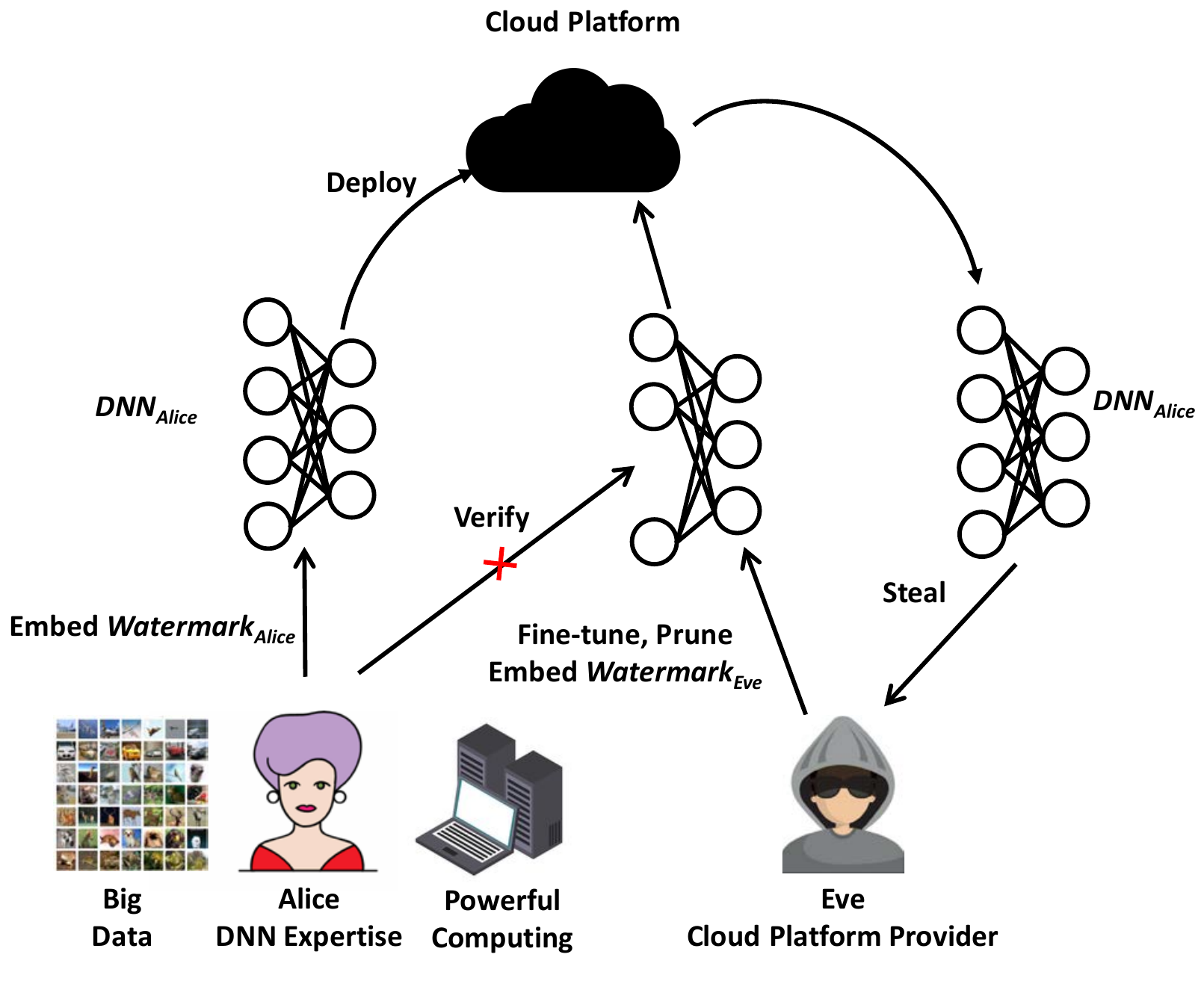, width=0.44\textwidth} 
\caption{A Motivating Example}
\label{fig:motivation}
\vspace{-5mm}
\end{figure}

With the advancement of AI techniques, neural networks are involved in various complicated tasks such as autonomous driving, speech recognition, medical diagnosis, etc. Such tasks typically rely on DNN models that are trained on a large amount of data using intensive GPU resources. For instance, the training of BERT~\cite{devlin2018bert}, a language understanding model, took 3.3 days (79.2 hours) on 4 DGX-2H servers with 64 Tesla V100 GPUs, and received \$3,751-\$12,571 bills from a cloud computing platform~\cite{strubell2019energy}. Moreover, it is estimated that training GPT-3, an autoregressive language model, would cost around \$12 million~\cite{GPT3}. Hence, protecting the IP of model owners is on demand. However, the existing backdoor-based watermarking approaches provide no sufficient protection for model owners~\cite{adi2018turning,zhang2018protecting,guo2018watermarking,rouhani2018deepsigns,namba2019robust,li2019prove}. Consider a motivating example as shown in Figure~\ref{fig:motivation}, where Alice develops a valuable DNN model $DNN_{Alice}$. Realizing the necessity of IP protection, Alice embeds several backdoor samples (i.e., triggers and their desired labels) into the model as a digital watermark $watermark_{Alice}$. Then, Alice deploys $DNN_{Alice}$ on a cloud platform in operation. 

Now assume that an attacker Eve (e.g., the cloud platform provider, an insider, or a hacker) who has access to the structure and parameters of $DNN_{Alice}$ steals the model and re-commercializes it for profit. To evade Alice's detection, Eve fine-tunes and/or prunes $DNN_{Alice}$, aiming to remove Alice's watermark. To prepare for possible ownership arguments, Eve also injects a number of backdoor samples of her own choice into the stolen model as her own digital watermark $watermark_{Eve}$\footnote{Alternatively, Eve can also generate a number of adversarial examples on the stolen model as her own digital watermark $watermark_{Eve}$.}. The question is, how Alice can testify that the model $DNN_{Eve}$ is stolen from her model $DNN_{Alice}$. Alice may present her set of triggers as inputs to $DNN_{Eve}$ and observe the model's output. Even if we assume that Eve's fine-tuning and/or pruning is not effective in removing Alice's watermark, Eve can still confuse ownership verification by presenting her set of triggers or adversarial examples, thus claiming her ownership of $DNN_{Eve}$. In the worst case, the pruning and/or fine-tuning of $DNN_{Alice}$ performed by Eve removes Alice's watermark from $DNN_{Eve}$. Consequently, Eve is victorious over Alice in the dispute of ownership. 

One may argue that Alice may increase the number of embedded triggers, and use it to outperform Eve. However, such approach is not feasible. On the one hand, with the knowledge that more triggers win, Eve can simply compete with Alice by embedding more backdoors as her own watermark. On the other hand, as the number of embedded triggers increases, the accuracy of the watermarked model decreases as shown in Table~\ref{tab:numberofbackdoor} in Appendix, which limits the number of backdoors one can embed into a model. We use the backdoor injection approach described in \cite{adi2018turning} to embed different number of backdoors into VGG11 (trained on CIFAR-10), and find the accuracy drops by 1.4\% with 400 backdoor samples embedded. In addition, we obtained code from authors of \cite{adi2018turning}, applied the same setting, trained Resnet18 on CIFAR-10, repeated their experiments and got similar results as in their paper. Then we embedded a different number of backdoors into the above model, and find the accuracy drops 1.68\% with 400 backdoor samples embedded. Therefore, Alice may not want to embed too many backdoor samples as it may compromise the accuracy of the watermarked model.

\subsection{Threat Model}
\label{subsec:threat}
We consider white-box adversaries who have full access to model structure and parameters. We assume that the adversaries, such as prowlers of neural networks, are proficient in machine learning and watermarking. To infringe on the IP of a watermarked DNN model, the adversaries may apply various machine learning techniques such as model pruning, model fine-tuning, kernels cutoff/supplement and crafting adversarial samples. Note that kernels cutoff or supplement cannot work for LSTM, since such operations will change the structure of the model, but LSTM is one piece of functional unit and its internal structure cannot be changed. Moreover, the adversaries are also capable of launching functionality-equivalent attack through structure adjustment or parameter adjustment to fail embedded watermarks. Note that such attack is only effective against CNN models for making the structure adjustment and against ReLU activation function for making the parameter adjustment. It does not work on LSTM, since it uses sigmoid as the activation function (making the parameter adjustment impossible) and the weights in LSTM must be in order (making the structure adjustment impossible). We assume that the adversaries lack necessary resources, including large-scale training datasets and sufficient computing resources, to retrain the entire or a large portion of stolen model. Finally, we also assume adversaries do not make changes beyond the functionality-equivalent and kernels cutoff attacks to the structure, which may reduce the accuracy of the stolen model to an unacceptable level.

\vspace{-4pt}
\section{Approach}
\label{sec:Approach}
\begin{figure*}[!t]
\centering
\epsfig{figure=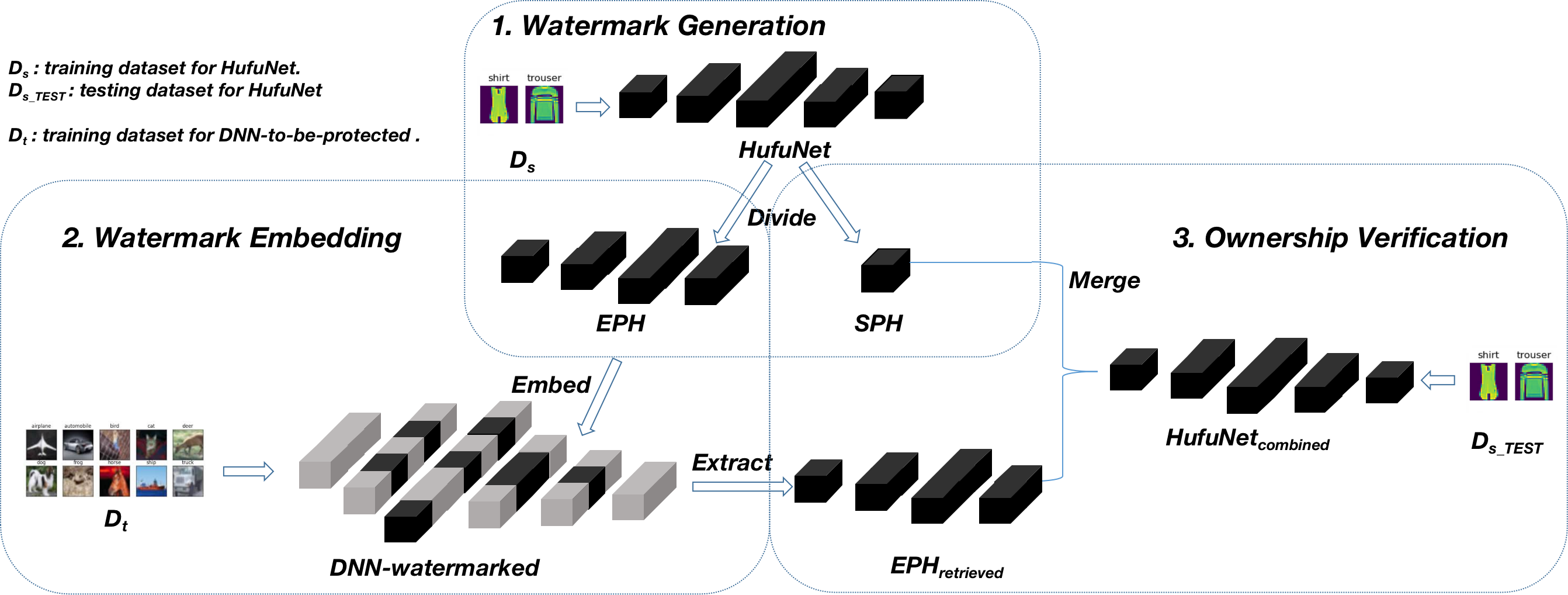, width=0.80\textwidth} 
\caption{Workflow of HufuNet}
\label{fig:workflow}
\end{figure*}

Figure~\ref{fig:workflow} shows the workflow of HufuNet in three phases: watermark generation, watermark embedding, and ownership verification. During the watermark generation phase, a model owner trains a unique neural network, HufuNet, on a public training set $D_s$ and divides its parameter space into two pieces: $Embedded\ Piece\ of\ HufuNet\ (EPH)$ as watermark, and $Secret\ Piece\ of\ HufuNet\ (SPH)$ as local secret. During the watermark embedding phase, the model owner injects the watermark EPH into the parameter space of his/her DNN model (i.e., DNN-to-be-protected) for ownership protection and trains it on a large training set $D_t$ with EPH frozen to fulfill its training task (i.e., DNN-watermarked). During the ownership verification phase, the owner extracts a watermark $EPH_{retrieved}$ from a suspect DNN model, and merges it with his/her local SPH to generate a neural network model $HufuNet_{combined}$. The owner can then test $HufuNet_{combined}$ on a public testing dataset $D_{s\_TEST}$ with the same distribution as $D_s$, and verify his/her ownership if the suspect model is embedded with the watermark EPH. Following Kerckhoffs' principle, we assume the watermark embedding algorithm and the ownership verification algorithm are public. While the adversaries may forge HufuNet from the public training dataset $D_s$, they need to use the public testing dataset $D_{s\_TEST}$ and follow the public ownership verification algorithm to claim their ownership over disputed models. 
\vspace{-3pt}
\subsection{Watermark Generation}
\label{subsec: watermark_Genaration}
The structure of HufuNet consists of several convolutional layers followed by a fully connected layer. The size of each convolution kernel in HufuNet is chosen to be $3*3$ in this paper; therefore, the number of parameters for each kernel is nine. After HufuNet is trained on a public data set $D_s$, all of its convolution layers are used as watermark EPH, while the fully connected layer is kept locally as secret SPH. The size of EPH is typically much smaller than the DNN model in which EPH is embedded. 
HufuNet can be considered as a mapping from its input space to its output space: $\mathbb{R}^M_{s} \rightarrow \mathbb{R}^N_{s}$ after it is trained using the dataset $D_{s}$.
The mapping achieves high accuracy on a testing dataset $D_{s\_TEST}$, which has the same distribution as $D_s$. Both $D_s$ and $D_{s\_TEST}$ are released to the public for training and verifying HufuNet. Note that it is beneficial to standardize $D_s$ and $D_{s\_TEST}$ so that the adversaries cannot use arbitrary datasets to forge and verify their own HufuNet (refer to evaluation results in Section~\ref{subsec:security1}).  

\vspace{-3pt}
\subsection{Watermark Embedding}
\label{subsec:Watermark Embedding}

We focus on CNN models in explaining HufuNet watermark embedding process, where the DNN-to-be-protected includes convolutional layers. Regarding RNN models, we simply view LSTM as one convolutional layer with $N$ kernels, each with $3*3$ parameters, to embed HufuNet watermarks.

\noindent\textbf{Embedding approach.}
Watermark EPH is embedded into the convolutional layers of DNN-to-be-protected. In this process, each convolution kernel of EPH is embedded individually into DNN-to-be-protected, which is much larger than EPH. A secret key determines the embedding location of each kernel based on its parameter values and index, to ensure the stealthiness of EPH in the parameter space of DNN-to-be-protected. The secret key should be strong enough to defend against potential attacks to detect the embedded watermark. e.g., brute force searching (refer to evaluation results in Section \ref{exp_stealthiness}).

Algorithm \ref{alg:Position-injected} shows the details of our watermark embedding approach. Line 1 initializes the variables $bitmap$, which keeps track of whether each position in DNN-to-be-protected has been embedded, and DNN-watermarked $f_{wm}$. For each convolution kernel in our watermark EPH (Line 2), we first perform an XOR operation on its index $i$ and all its $k*k$ parameter ($k=3$ in our EPH), hash the result using a secret $key$, and finally do a modulo over the number of kernels in DNN-to-be-protected to compute the embedding position (Line 3). If the position has been embedded, we simply do a linear probing to find the next available position (Line 4-6). Once the position is found, we embed the convolution kernel into the corresponding position in $f_t$ (Line 7), update $bitmap$ to indicate the position has been used (Line 8), and continue the next kernel (Line 9). Finally, we output DNN-watermarked $f_{wm}$ in Line 11. Note that the kernel parameters in EPH are used in determining where EPH is embedded in $f_t$. The owner of $f_t$ needs to keep the whole HufuNet, including EPH and SPH, for ownership verification. In the process of ownership verification, the preserved EPH is used to locate where it is embedded from a suspect model, while the values in the suspect model are not used to locate the watermark since they may have been changed in the watermark embedding process and possibly by the adversaries. 

\begin{algorithm}[t]
 	\caption{Embedding Algorithm}
 	\label{alg:Position-injected}
 	\begin{algorithmic}[1]
 	\REQUIRE $f_t$: DNN-to-be-protected; $N$: number of convolution kernels in $f_t$; $Kernel$: all kernels in EPH to be embedded;
 	 $n$: number of kernels in $Kernel$; $bitmap$: a bitmap to keep track of the usage of each position in $f_t$; $key$: a secret key 
 	\ENSURE $f_{wm}$: DNN-watermarked 
 	\STATE{$bitmap = NULL$, $f_{wm} = f_t$}
	\FOR{$i$ in $(1, n)$}
        \STATE{$position=HMAC(key,XORPMV(kernel_i) \oplus i) \% N$}
		\WHILE{$bitmap_{position}$} 
		\STATE{$position = ++position \% $ N}
		\ENDWHILE
		\STATE{$f_{wm} = EMBED(kererl_i, position, f_{wm}$)}
		\STATE{$bitmap_{position} = 1$}
		\STATE{$++i$}
	\ENDFOR
    \RETURN $f_{wm}$
\end{algorithmic}
\rule[0pt]{8.4cm}{0.05em}
\begin{tablenotes}
\footnotesize
\item{$XORPMV$ is the function to $XOR$ all the parameter values of a convolution kernel. $EMBED$ is the function to embed a convolution kernel of EPH into a specific position of $f_t$. $HMAC$ is a keyed cryptographic hash function.}
\end{tablenotes}
\end{algorithm}

\noindent\textbf{Training DNN-watermarked.} 
The DNN-to-be-protected $f_t$ is a mapping from its input space $\mathbb{R}^M_t$ to its output space $\mathbb{R}^N_t$. Before training, the owner of $f_t$ randomly initializes the parameters of $f_t$, and embeds the parameters of EPH based on the approach discussed above. The owner then trains $f_t$ on its training dataset $D_t$. During the training process of $f_t$, the owner freezes the parameter space of EPH and only updates the rest of the parameter space. In this way, the parameter space of EPH remains unchanged while other parameters of $f_t$ are updated during training. Note that the loss function of $f_t$ remains unchanged during the training process.

\vspace{-3pt}
\subsection{Ownership Verification}
\label{subsec:Ownership Verification}
Given a model $f_{suspect}$, the owner first extracts the parameter space $EPH_{retrieved}$ from $f_{suspect}$. A straightforward extraction method is to repeat Algorithm \ref{alg:Position-injected} in Section \ref{subsec:Watermark Embedding}. Based on the original HufuNet kept by the owner, he/she can compute the locations in $f_{suspect}$ where EPH is supposedly embedded, and retrieve the corresponding parameter space as $EPH_{retrieved}$. Then the owner merges $EPH_{retrieved}$ with the preserved SPH to obtain $HufuNet_{combined}$. If $f_{suspect}$ is stolen from $f_{wm}$, the rebuilt $HufuNet_{combined}$ is expected to maintain a high accuracy on the standard test set $D_{s\_TEST}$, which has the same distribution as the standard training dataset $D_{s}$. Specifically, the owner evaluates the accuracy $Acc_{combined}$ of $HufuNet_{combined}$ using $D_{s\_TEST}$. If the adversary does not retrain $f_{suspect}$, $Acc_{combined}$ will have the same value as the accuracy of the original HufuNet (i.e., $Acc_{ori}$). Otherwise, $Acc_{combined}$ would drop. To solve this problem, we compare $Acc_{combined}$ with $Acc_{ori}$ using $Diff_{Acc} = |Acc_{ori} - Acc_{combined}|/Acc_{ori}$, if $Diff_{Acc} < \tau_{acc}$, we think that $f_{suspect}$ infringes the IP of the DNN-watermarked $f_{wm}$. $\tau_{acc}$ is the threshold controlling the difference between $Acc_{ori}$ and $Acc_{combined}$. In our evaluation, we find that HufuNet is highly robust against model pruning, fine-tuning and kernels cutoff. Taking CNN as an example, even if the accuracy of $f_{suspect}$ drops by 50.74\%, $Acc_{combined}$ can still be higher than 80\% and $Diff_{Acc}$ is 13.34\% (details are in Section~\ref{subsec:Robustness}). Therefore, we define $\tau_{acc}$ as 15\% in our evaluation, which is good enough to preserve good robustness and integrity of HufuNet.

\subsection{Restore before Retrieve}
\label{subsec:restore}
As a parameter level watermark, HufuNet may suffer from the functionality-equivalent attacks as discussed in Section \ref{sub:destructing}. Therefore, even if $Diff_{Acc}$ is larger than $\tau_{acc}$, a model can still be suspicious ($f_{suspect}$) if it has exactly the same or quite similar structure as the protected model, e.g., the same number of layers and the same number of parameters (with different order) on each layer, or the same number of layers and the same number of kernels on most layers. In such scenarios, we examine $f_{suspect}$ further using the following restore approach and determine whether or not such model has been intentionally adjusted to evade our parameter level watermark. Note that if $f_{suspect}$ is an innocent model, the restore approach will not falsely claim ownership over it.

Regarding structure adjustment, we recover the order of input/output channels of $f_{suspect}$ according to the locally kept HufuNet $f_{wm}$. We start from the first layer where only the order of the output channels can be changed. We compare the parameters of each output channel of $f_{suspect}$ with those of $f_{wm}$ using cosine similarity, so as to find the optimal match.
In this way, we can recover the order of each output channel of the first layer in $f_{suspect}$ according to  that of our locally stored $f_{wm}$. Then the input channels of the second layer can be restored straightforwardly based on the restored order of the output channels of the first layer. Similarly, the output/input channels of all other layers can be restored accordingly based on the locally kept $f_{wm}$, thus enabling correct retrieval of $EPH_{retrieved}$ even if $f_{suspect}$ experienced structure adjustment. The above cosine similarity approach also works well when many parameters in output channels are adjusted, such as fine-tuning (detailed discussion in Appendix \ref{analysis_of_channel_expansion_principle}).

Regarding parameter adjustment, we first ensure that the order of the input/output channels of $f_{suspect}$ is recovered using the approach described above. Then we perform Singular Value Decomposition (SVD) on each convolution kernel of $f_{suspect}$ and $f_{wm}$ to obtain intermediate singular value matrices. For each pair of singular value matrix in $f_{suspect}$ and singular value matrix in $f_{wm}$, we divide each element of the former by each element of the latter; thus we get a series of resultant matrices. The values on the diagonal of each resultant matrix are averaged to compute the scale factor $c$, used to obtain the restored convolution kernel (i.e., dividing each element in the corresponding convolution kernel $f_{suspect}$ by $c$). We utilize this method to restore all the convolution kernels.

Note that the scale factor is not computed by (i) dividing each element of the convolution kernel in $f_{suspect}$ by each element of the convolution kernel in our local $f_{wm}$, and (ii) computing the average value of the resultant matrix. This is because the singular values of a convolution kernel represent the scaling characteristics of the kernel; it is more reliable to compute the scaling factor from the singular values rather than from kernel parameters directly. This is clear in the case where an attacker first fine-tunes a stolen model and then perform the functionality-equivalent attack through parameter adjustment. The fine-tuning typically makes those smaller values in the convolution kernel change more significantly than those larger ones (e.g., increasing from 1 to 2 results in the factor of 2, but increasing from 10 to 11 yields the factor of 1.1).

Meanwhile, a highly suspicious model $f_{suspect}$ may show minor structural differences, e.g., the number of kernels on one or several layers of $f_{suspect}$ is smaller or larger than those of our watermarked model. We further examine whether or not some kernels of such model have been intentionally cut off or supplemented to evade our watermark. 
First, we still restore the structure of $f_{suspect}$ using cosine similarity approach presented above, since kernels cutoff/supplement attacks always change the order of kernels as well, unless all the cutoff/supplemented kernels are the last ones of the layer. With all the kernels of $f_{suspect}$ restored and our locally stored $f_{wm}$ as a reference, we fill in missing (cutoff) kernels with all parameters as 0 or cut off extra (supplemented) kernels directly. Note that using the values of the corresponding kernels in $f_{wm}$ to fill in may cause false positive on innocent models.
In this way, we can restore $f_{suspect}$ with the same structure as $f_{wm}$ and the same number of kernels $N$, so Algorithm \ref{alg:Position-injected} can be used to extract EPH.

\vspace{-4pt}
\section{Evaluation}
\label{sec:Evaluation}

Based on the watermark destruction approaches discussed in Section \ref{sub:destructing}, we evaluate our solution in the following aspects. (i) robustness: watermarks should still be detected by owners from stolen DNN models even if the models experienced fine-tuning, pruning and kernels cutoff/supplement (refer to Section \ref{subsec:Robustness}); (ii) adaptiveness: watermarks should still be detected by owners even if the model has been converted to a functionality-equivalent one via structure adjustment, parameter adjustment or channel expansion (refer to Section \ref{subsec:adaptiveness}); (iii) against synthetic attacks: watermark can still be detected by owners even if the attackers applied fine-tuning, pruning and functionality-equivalent attacks together on their stolen models (refer to Section \ref{subsec:combination adaptiveness with robustness}); (iv) stealthiness: it is difficult for attackers to learn the existence of watermark from a stolen DNN model (refer to Section \ref{exp_stealthiness}); (v) security: it is difficult for attackers to forge a valid watermark for a stolen DNN model (also called ambiguity attack~\cite{fan2019rethinking}, refer to Section \ref{subsec:security1}); (vi) integrity: it is highly unlikely for DNN owners to claim ownership over innocent DNN models (refer to Section \ref{subsec:integrity}); and (vii) fidelity: watermark-embedding should impact little on the performance of the original DNN models (refer to Section \ref{subsec:fidelity}).

\subsection{Experimental Setup}
\label{subsec:Experiment Setting}

\noindent\textbf{Models}. Without loss of generality, we conduct experiments on five models: VGG11~\cite{simonyan2014very}, GoogLeNet~\cite{szegedy2015going}, Resnet18 and Resnet34~\cite{he2016deep}, and LSTM~\cite{hochreiter1997long}, which are all used as DNNs-to-be-protected. HufuNet is designed with five convolutional layers (used as EPH) and one fully connected layer (used as SPH).


\noindent\textbf{Datasets}. We choose four datasets: Fashion-MNIST~\cite{xiao2017fashion}, IMDB~\cite{maas-EtAl:2011:ACL-HLT2011}, CIFAR-10, and CIFAR-100 ~\cite{krizhevsky2009learning} for the above five models accordingly. 
We use Fashion-MNIST to train our HufuNet, CIFAR-10 to train VGG11, GoogLeNet as well as Resnet18, CIFAR-100 to train Resnet34, and IMDB to train LSTM as shown in Table \ref{tab:datasets} in Appendix. The classification task and the main task of the DNN models are used interchangeably thereafter.

\noindent\textbf{Platform}. All our experiments are conducted on a server running 64-bit Ubuntu 18.04 system with Intel Xeon E5-2620 v4 @ 2.10GHz CPU, 128GB memory, 3TB hard drive and 3 Nvidia GPU TiTan X GPUs each with 12GB memory.


\vspace{-8pt}
\subsection{Baseline Performance}
\label{subsec:Performance}
We evaluate the ownership verification of HufuNet on the model without any changes between embedding and verifying, i.e., assuming adversaries just steal the model without performing any other operations on the stolen model, as the baseline performance. The number of parameters of the five DNNs-to-be-protected and HufuNet are shown in Table \ref{tab:parameter} in Appendix. It can be seen that our watermark EPH only consumes a small portion of the parameter space of VGG11, GoogLeNet, Restnet18 and Resnet34, ranging from {1.48}\% to {5.37}\%), since the number of parameters of these four models are relatively large. Since LSTM is intentionally selected to evaluate the performance of HufuNet on smaller models, EPH consumes up to 13.38\% of its parameter space, considering LSTM only contains 9,251 KB parameters in just three layers.

\begin{table}
\begin{threeparttable}
\centering
\footnotesize
\caption{Baseline Performance of Ownership Verification}
\label{tab:Performance of ownership verification}
\begin{tabular}{m{1.0cm}
<{\centering}|m{0.8cm}
<{\centering}|m{1.3cm}
<{\centering}|m{1.0cm}
<{\centering}|m{1.0cm}
<{\centering}|m{0.7cm}
<{\centering}}
\hline
\textbf{ }& \textbf{VGG11}& \textbf{GoogLeNet}& \textbf{Resnet18}& \textbf{Resnet34}& \textbf{LSTM}\\
\hline \hline
\textbf{Original$^{1}$} & \text{77.77$\%$} &\text{90.85$\%$}& \text{86.14$\%$}&\text{70.20$\%$}& \text{87.08$\%$} \\ \hline
\textbf{Suspect$^{2}$} & \text{77.77$\%$} &\text{90.85$\%$}& \text{86.14$\%$}&\text{70.20$\%$}& \text{87.08$\%$} \\ \hline
\textbf{$\tau_{acc}^{3}$}& \text{15$\%$} &\text{15$\%$}& \text{15$\%$}&\text{15$\%$}& \text{15$\%$} \\ \hline
\textbf{Combine$^{4}$}& \text{92.32$\%$} &\text{92.32$\%$}& \text{92.32$\%$}&\text{92.32$\%$}& \text{92.32$\%$} \\ \hline
\end{tabular}
\begin{tablenotes}
\item[1] Original indicates the accuracy of the original DNN-watermarked. 
\item[2] Suspect indicates the accuracy of the suspect model we would like examine. Since there is not any kind of retraining conducted, it demonstrates the same accuracy as that of our DNN-watermarked. 
\item[3] The threshold $\tau_{acc}$ is set as 15\%, so $HufuNet_{combined}$ accuracy above 80\% indicates ownership verification according to Section \ref{subsec:Ownership Verification}.
\item[4] Combine indicates the accuracy of the $HufuNet_{combined}$. Since there is not any kind of retraining conducted, it demonstrates the same accuracy as that of our original HufuNet. 
\end{tablenotes}
\end{threeparttable}
\end{table}


\begin{table*}[h]
\centering
\footnotesize
\caption{Robustness against Fine-tuning}
\label{tab:Robustness against Finetuning}
\begin{tabular}{m{0.7cm}
<{\centering}|m{1.42cm}
<{\centering}|m{1.0cm}
<{\centering}|m{1.42cm}
<{\centering}|m{1.0cm}
<{\centering}|m{1.42cm}
<{\centering}|m{1.0cm}
<{\centering}|m{1.42cm}
<{\centering}|m{1.0cm}
<{\centering}|m{1.42cm}
<{\centering}|m{1.0cm}
<{\centering}}

\hline
\multirow{2}{*}{\textbf{\makecell[c]{Epochs}}} & \multicolumn{2}{c|}{\textbf{VGG11}}& \multicolumn{2}{c|}{\textbf{GoogLeNet}}& \multicolumn{2}{c|}{\textbf{Resnet18}}&\multicolumn{2}{c|}{\textbf{Resnet34}}& \multicolumn{2}{c}{\textbf{LSTM}}\\ \cline{2-11}
\text{}&\textbf{DNN-watermarked}&\textbf{HufuNet}&\textbf{DNN-watermarked}& \textbf{HufuNet}&\textbf{DNN-watermarked}&\textbf{HufuNet}& \textbf{DNN-watermarked}&\textbf{HufuNet}& \textbf{DNN-watermarked}&\textbf{HufuNet}\\

\hline \hline
\text{0}&\text{77.77$\%$}&\text{92.32$\%$}&\text{$90.85\%$}& \text{$92.32\%$}&\text{$86.14\%$}&\text{$92.32\%$}&\text{$70.20\%$}&\text{$92.32\%$}& \text{$87.08\%$}&\text{$92.32\%$}\\

\hline
\text{10}&\text{75.35$\%$}&\text{92.37$\%$}&\text{$89.55\%$}& \text{$91.91\%$}&\text{$85.75\%$}&\text{$92.34\%$}& \text{$66.97\%$}&\text{$92.26\%$}& \text{$85.90\%$}&\text{$92.28\%$}\\

\hline 
\text{20}&\text{74.70$\%$}&\text{92.20$\%$}&\text{$89.75\%$}& \text{$91.99\%$}&\text{$85.70\%$}&\text{$92.35\%$}&\text{$66.78\%$}&\text{$92.20\%$}& \text{$86.10\%$}&\text{$92.36\%$}\\

\hline 
\text{30}&\text{72.30$\%$}&\text{92.21$\%$}&\text{$89.75\%$}& \text{$92.03\%$}&\text{$85.60\%$}&\text{$92.36\%$}& \text{$68.19\%$}&\text{$92.15\%$}&\text{$85.60\%$}&\text{$92.18\%$}\\

\hline 
\text{40}&\text{73.65$\%$}&\text{92.16$\%$}&\text{$89.70\%$}& \text{$92.07\%$}&\text{$85.65\%$}&\text{$92.36\%$}& \text{$67.44\%$}&\text{$92.01\%$}& \text{$88.20\%$}&\text{$92.19\%$}\\

\hline 
\text{50}&\text{73.35$\%$}&\text{92.19$\%$}&\text{$89.70\%$}& \text{$92.06\%$}&\text{$85.50\%$}&\text{$92.32\%$}&\text{$67.48\%$}&\text{$91.83\%$}&  \text{$87.80\%$}&\text{$92.22\%$}\\

\hline
\text{60}&\text{75.80$\%$}&\text{92.18$\%$}&\text{$89.80\%$}& \text{$92.04\%$}&\text{$85.50\%$}&\text{$92.39\%$}& \text{$67.64\%$}&\text{$91.88\%$}&\text{$88.00\%$}&\text{$92.17\%$}\\

\hline 
\text{70}&\text{75.80$\%$}&\text{92.16$\%$}&\text{$89.70\%$}& \text{$92.02\%$}&\text{$85.55\%$}&\text{$92.37\%$}&\text{$67.51\%$}&\text{$91.85\%$}& \text{$87.30\%$}&\text{$92.11\%$}\\

\hline 
\text{80}&\text{75.85$\%$}&\text{92.16$\%$}&\text{$89.65\%$}& \text{$92.01\%$}&\text{$85.70\%$}&\text{$92.39\%$}&\text{$67.81\%$}&\text{$91.53\%$}& \text{$87.50\%$}&\text{$92.14\%$}\\

\hline
\text{90}&\text{75.70$\%$}&\text{92.14$\%$}&\text{$89.45\%$}& \text{$92.03\%$}&\text{$85.70\%$}&\text{$92.39\%$}& \text{$66.91\%$}&\text{$91.32\%$}&\text{$87.50\%$}&\text{$92.16\%$}\\

\hline
\text{100}&\text{75.60$\%$}&\text{92.14$\%$}&\text{$89.45\%$}& \text{$92.03\%$}&\text{$85.70\%$}&\text{$92.39\%$}&\text{$67.67\%$}&\text{$91.08\%$}& \text{$87.60\%$}&\text{$92.16\%$}\\

\hline
\end{tabular}
\begin{tablenotes}
\item \text{The HufuNet in the table indicates the accuracy of $HufuNet_{combined}$.}
\end{tablenotes}
\end{table*}

\begin{table*}[h]
\centering
\footnotesize
\caption{Robustness against Pruning}
\label{Robustness against Pruning}
\begin{tabular}{m{0.7cm}
<{\centering}|m{1.42cm}
<{\centering}|m{1.0cm}
<{\centering}|m{1.42cm}
<{\centering}|m{1.0cm}
<{\centering}|m{1.42cm}
<{\centering}|m{1.0cm}
<{\centering}|m{1.42cm}
<{\centering}|m{1.0cm}
<{\centering}|m{1.42cm}
<{\centering}|m{1.0cm}
<{\centering}}
\hline
\multirow{2}{*}{\textbf{\makecell[c]{pct.}}} & \multicolumn{2}{c|}{\textbf{VGG11}}& \multicolumn{2}{c|}{\textbf{GoogLeNet}}& \multicolumn{2}{c|}{\textbf{Resnet18}}&\multicolumn{2}{c|}{\textbf{Resnet34}}& \multicolumn{2}{c}{\textbf{LSTM}}\\ \cline{2-11}
\text{}&\textbf{DNN-watermarked}&\textbf{HufuNet}&\textbf{DNN-watermarked}&\textbf{HufuNet}&\textbf{DNN-watermarked}&\textbf{HufuNet}& \textbf{DNN-watermarked}&\textbf{HufuNet}&\textbf{DNN-watermarked}&\textbf{HufuNet}\\

\hline \hline
\text{0}&\text{77.77$\%$}&\text{92.32$\%$}&\text{$90.85\%$}& \text{$92.32\%$}&\text{$86.14\%$}&\text{$92.32\%$}&\text{$70.20\%$}&\text{$92.32\%$}& \text{$87.08\%$}&\text{$92.32\%$}\\

\hline
\text{10$\%$}&\text{77.68$\%$}&\text{92.33$\%$}&\text{$90.77\%$}& \text{$92.29\%$}&\text{$86.14\%$}&\text{$92.29\%$}&\text{$70.29\%$}&\text{$92.27\%$}& \text{$87.06\%$}&\text{$92.33\%$}\\

\hline 
\text{$20\%$}&\text{77.80$\%$}&\text{92.29$\%$}&\text{$90.55\%$}& \textbf{$92.32\%$}&\text{$85.96\%$}&\text{$92.34\%$}&\text{$70.20\%$}&\text{$92.46\%$}& \text{$87.18\%$}&\text{$92.37\%$}\\

\hline 
\text{$30\%$}&\text{77.82$\%$}&\text{92.27$\%$}&\text{$90.19\%$}& \text{$92.32\%$}&\text{$85.75\%$}&\text{$92.35\%$}&\text{$69.70\%$}&\text{$92.20\%$}& \text{$87.26\%$}&\text{$92.36\%$}\\

\hline 
\text{40$\%$}&\text{$77.64\%$}&\text{92.37$\%$}&\text{$85.90\%$}& \text{$92.36\%$}&\text{$82.24\%$}&\text{$92.37\%$}&\text{$67.25\%$}&\text{$91.98\%$}& \text{$86.94\%$}&\text{$92.19\%$}\\

\hline 
\text{50$\%$}&\text{77.23$\%$}&\text{92.29$\%$}&\text{$56.46\%$}& \text{$92.07\%$}&\text{$74.81\%$}&\text{$92.09\%$}&\text{$56.46\%$}&\text{$91.21\%$}& \text{$86.85\%$}&\text{$91.86\%$}\\

\hline
\text{60$\%$}&\text{75.68$\%$}&\text{92.25$\%$}&\text{$19.63\%$}& \text{$91.73\%$}&\text{$46.84\%$}&\text{$91.74\%$}&\text{$27.86\%$}&\text{$89.19\%$}& \text{$86.89\%$}&\text{$91.36\%$}\\

\hline 
\text{70$\%$}&\text{69.84$\%$}&\text{91.86$\%$}&\text{$11.73\%$}& \text{$90.29\%$}&\text{$11.44\%$}&\text{$90.35\%$}&\text{$4.86\%$}&\text{$85.60\%$}& \text{$86.28\%$}&\text{$89.95\%$}\\

\hline 
\text{$80\%$}&\text{50.74$\%$}&\text{90.99$\%$}&\text{$10.00\%$}& \text{$87.63\%$}&\text{$10.00\%$}&\text{$87.80\%$}&\text{$1.10\%$}&\text{$65.61\%$}& \text{$84.16\%$}&\text{$85.74\%$}\\

\hline
\text{$90\%$}&\text{21.74$\%$}&\text{87.54$\%$}&\text{$10.00\%$}& \text{$66.12\%$}&\text{$10.00\%$}&\text{$67.83\%$}&\text{$1.00\%$}&\text{$34.82\%$}& \text{$60.75\%$}&\text{$58.14\%$}\\

\hline
\end{tabular}
\begin{tablenotes}
\footnotesize
\item[*]\text{pct. indicates the percentage of pruning. The HufuNet in the table indicates the accuracy of $HufuNet_{combined}$.}
\end{tablenotes}
\end{table*}

The experimental results are shown in Table~\ref{tab:Performance of ownership verification}. Since there is no retraining conducted on the original DNN-watermarked, the suspect model (to-be-examined by HufuNet for ownership verification) demonstrates the same accuracy as that of our original model. Similarly, $EPH_{retrieved}$ is exactly the same as original EPH, so the accuracy of $HufuNet_{combined}$ equals to that of the original HufuNet. Therefore, the rebuilt $HufuNet_{combined}$ can always verify the ownership on the suspect model accurately considering not any kind of retraining on the stolen model is conducted\footnote{We also evaluate the integrity of HufuNet in Section \ref{subsec:integrity}, that is, without EPH, we will not claim ownership against innocent models.}.
According to Section \ref{subsec:Ownership Verification}, with $Acc_{ori}$ as 92.32\% and $\tau_{acc}$ as 15\%, the accuracy of $HufuNet_{combined}$ above 80\% means successful ownership verification. In the following experiments, we will just show HufuNet (combined) accuracy to indicate ownership verification.


\begin{table}
\centering
\footnotesize
\caption{Robustness against Kernels Cutoff}
\label{tab:robustnesscuttingoffkernels}
\begin{tabular}{m{1.17cm}
<{\centering}|m{1.5cm}
<{\centering}|m{2.5cm}
<{\centering}|m{1.2cm}
<{\centering}}
\hline
\textbf{DNN}&\textbf{Variance}& \textbf{Accuracy of DNN-watermarked}&\textbf{HufuNet Accuracy} \\
\hline \hline
\multirow{3}{*}{\text{VGG11}}&\text{Original}&\text{77.77$\%$}&\text{92.32$\%$} \\
\text{}&\text{Cut off}&\text{72.70$\%$}& \text{-} \\
\text{}&\text{Restore}&\text{72.70$\%$}& \text{92.14$\%$}\\ \hline

\multirow{3}{*}{\text{Resnet18}}&\text{Original}&\text{86.14$\%$}& \text{92.32$\%$} \\
\text{}&\text{Cut off}&\text{80.28$\%$}& \text{-} \\
\text{}&\text{Restore}&\text{80.28$\%$}& \text{92.38$\%$}\\ \hline

\multirow{3}{*}{\text{GoogLeNet}}&\text{Original}&\text{90.85$\%$}& \text{92.32$\%$} \\
\text{}&\text{Cut off}&\text{85.10$\%$}& \text{-} \\
\text{}&\text{Restore}&\text{85.10$\%$}& \text{92.02$\%$}\\ \hline

\end{tabular}
\end{table}

\vspace{-4pt}
\subsection{Robustness} 
\label{subsec:Robustness}
Due to the stealthiness of EPH (as shown in Section \ref{exp_stealthiness}), a targeted destruction (knowing where EPH is and destructing it correspondingly) is almost impossible. The adversary certainly can initialize and retrain all the convolutional layers of the stolen model. In this scenarios however, the adversary would rather train his/her own model than stealing $f_{wm}$, since those two procedures demand similar amount of computing resources and training data. In contrast, adversaries would prefer ``blindly'' revising the model, e.g., fine-tuning, pruning, kernels cutoff/supplement, etc., hoping to destruct embedded watermarks. 

\noindent\textbf{Robustness against fine-tuning.}
\label{finetunig}
We trained five models in total (with our watermark EPH embedded): VGG11, GoogLeNet, Resnet18 on CIFAR-10, Resnet34 on CIFAR-100, and LSTM on IMDB. We assume adversaries only have access to the testing dataset (provided by the model owner to test the performance of the model, including 10,000 samples), and try to leverage it to fine-tune the stolen model. Referring to the settings in \cite{li2019prove}, adversaries divide their available dataset into two parts, 80\% as the training dataset and 20\% as testing dataset. Same as \cite{adi2018turning}, we also set initial learning rate $\lambda=0.001$ and the delay factor as 0.0005 to fine-tune DNNs-watermarked using SGD (Stochastic Gradient Descent). The maximum number of epochs for fine-tuning is set as 100 according to \cite{li2019prove}, and the incremental step of the epoch is set as 10. As $\lambda$ decreases to a very small number, the parameters of DNNs-watermarked will rarely change, which indicates the model tends to be stable during fine-tuning. As shown in Table~\ref{tab:Robustness against Finetuning}, regarding the DNNs-watermarked, their accuracy drops at different scale when fine-tuned up to 40 epochs. As the number of epochs keeps increasing, the accuracy either gets improved or becomes stable. In contrast, the accuracy of $HufuNet_{combined}$ tends to be very stable, always above 90\%, which suffices ownership verification (above 80\%). Therefore, the fine-tuning process does not affect our EPH.

\noindent\textbf{Robustness against pruning.} 
To evaluate the robustness of HufuNet against pruning, we first embed our watermark EPH into VGG11, GoogLeNet, Resnet18, Resnet34 and LSTM, and then prune 10\% to 90\% parameters of each model based on their absolute values respectively. We adopt the commonly used pruning technique in ~\cite{han2015learning}, which eliminates the parameters with smaller absolute values, thus believed to have little impact on the performance of the models. We choose such model pruning approach, which is consistent with our assumption of the adversaries' capability, i.e., lack of sufficient training dataset and/or computing resources. Table~\ref{Robustness against Pruning} shows the experimental results on the accuracy of the corresponding DNNs-watermarked and HufuNet after pruning. Based on the results, we find that even if the DNNs-watermarked have been pruned to be considered as ``fail'' on the main task (e.g., the accuracy of VGG11, GoogLeNet, Resnet18, Resnet34 has been reduced to 50.74\%, 10.00\%, 10.00\% and 1.10\%, respectively) after 80\% pruning, the accuracy of $HufuNet_{combined}$ is still above 80\%, satisfying the preset threshold $\tau_{acc}$. For 90\% pruning, the accuracy of $HufuNet_{combined}$ drops below 80\%, but this high degree of pruning almost completely ruin the main task of the model. It is interesting that the accuracy of LSTM does not drop as significantly as other models, probably due to its larger parameter redundancy when dealing with simple tasks. Overall, the high accuracy of $HufuNet_{combined}$ upon heavy pruning indicates it is highly robust against pruning.

\noindent\textbf{Robustness against kernels cutoff/supplement.}
We evaluate the robustness of HufuNet against kernels cutoff attack on VGG11, Resnet18, and GoogLeNet. Resnet34 and LSTM are not evaluated, since the structure of the former is similar as Resnet18 and the latter is innately immune from such attack (as discussed in Section \ref{subsec:threat}). Since the removal of the convolution kernels will reduce the performance of $f_{cutoff}$, we control the kernels cutoff rate as 6.75\%, 2.30\% and 4.77\% respectively for the above three models, to ensure their accuracy does not drop significantly (around 5\%), i.e., 5.07\%, 5.86\% and 5.75\% respectively. Then we use the above proposed cosine similarity approach to restore the structure of $f_{cutoff}$ and supplement 0 for the missing kernels (cut off). As shown in Table \ref{tab:robustnesscuttingoffkernels}, the accuracy of HufuNet is 92.14\%, 92.38\%, and 92.02\% for the three models respectively, which are sufficient to verify the ownership (above 80\%). The evaluation result of kernels supplement is quite similar to that of kernels cutoff, since both the structure and the number of kernels are changed in a similar way. 

\noindent\textbf{Analysis of Robustness.} Due to the stealthiness (evaluated in Section \ref{exp_stealthiness}) and small scale of EPH in DNN-watermarked (EPH consumes 1.48\%$\sim$3.44\% of the parameter space of CNN models, and 13.38\% of the parameter space of the LSTM model, based on our experiment in Section \ref{subsec:Performance}), it is difficult for model pruning and fine-tuning to destroy the embedded watermark. The robustness of EPH against pruning and fine-tuning is also attributed to the parameter redundancy in DNN models. Usually a portion of parameters (e.g., the top 50\% parameters with large absolute values) contributes most to the task of a DNN model, while the other parameters can be considered ``redundant'' in the parameter space, making minor contribution to the model's task. Model retraining mainly affects those not-so-important values in the parameter space of DNN-watermarked; otherwise, the performance of the retrained model would be significantly worse than the watermarked model, which defeats the purpose of ownership verification. As long as a majority of significant parameters in EPH remains stable in model retraining, ownership verification can be performed robustly.

\vspace{-4pt}
\subsection{Adaptiveness}
\label{subsec:adaptiveness}
Since HufuNet is essentially a parameter-level watermark, we must ensure the embedded watermark can still be correctly retrieved to verify ownership. We further evaluate it against functionality-equivalent attack in terms of structure adjustment and parameter adjustment on three neural network models: VGG11, Resnet18, and GoogLeNet. We choose not to evaluate Resnet34 and LSTM, since the structure of the former is similar to Resnet18 and the latter is a RNN with sigmoid as activation function (which is not subject to functionality-equivalent attack).

\noindent\textbf{Adaptiveness against Structure Adjustment.}
To simulate the maximum capability of attackers, we shuffle the order of all convolution kernels of each layer, and at the same time we adjust the order of the inner layers of the convolution kernels to ensure functionality-equivalent. If there is a BatchNorm layer connected after a convolution layer, we also change its mean and variance according to the order of the corresponding convolution kernels. We use the approach discussed in Section \ref{subsec:restore} to restore the order of kernels. The experimental results are shown in Table \ref{tab:structureequivalent}. After reordering the convolution kernels of the DNNs-watermarked, the accuracy of VGG11, Resnet18 and GoogLeNet remains the same (functionally equivalent), but the accuracy of the combined HufuNet (without restore) decreases significantly from $92.32\%$ to $18.15\%$, $13.71\%$ and $8.22\%$ respectively. After restoring however, the accuracy of the combined HufuNet resumes to $92.32\%$, the same as the original HufuNet, thus allowing correct ownership verification. 

\noindent\textbf{Adaptiveness against Parameter Adjustment.} We also evaluate HufuNet against functionality-equivalent attack through parameter adjustment. 
We change the parameter values of each layer by multiplying these parameters by $c = 2^n$ (n is a randomly-choose integer for each layer ranging from -18 to 18), while still maintaining the functionality-equivalent of the overall model. We use the approach as discussed in Section \ref{subsec:restore} to restore the parameter values. The experiment result are shown in Table \ref{tab:parameterequivalent}. After adjusting parameters of the DNNs-watermarked (PMCG, i.e., parameter change), the accuracy of VGG11, Resnet18 and GoogLeNet remains the same (functionally equivalent), but the accuracy of the combined HufuNet (without restore) decreases significantly to $14.68\%$, $15.79\%$ and $25.52\%$ respectively. After restore however, the accuracy of the combined HufuNet resumes to $92.32\%$, the same as the original HufuNet, thus allowing ownership verification. 

\noindent\textbf{Adaptiveness against Channel Expansion.} \label{evaluation_channel_expansion} 
We also evaluate HufuNet against channel expansion on VGG11. We randomly choose two adjacent layers (the fourth and fifth layers), and double the output dimension of the fifth layer and input dimension of the sixth layer by supplementing output/input channels (details in Appendix \ref{analysis_of_channel_expansion_principle}).
Finally, we launch the structure adjustment attack to make it more difficult to trace the EPH. We use the approach discussed in Section \ref{subsec:restore} to restore EPH. After channel expansion and structure adjustment, the accuracy of VGG11 remains the same (functionally equivalent), but the accuracy of the combined HufuNet (without restore) drops to 9.91\%. After restore however, the accuracy of the combined HufuNet resumes to $92.32\%$, the same as the original HufuNet, thus allowing ownership verification.

\begin{table}
\centering
\footnotesize
\caption{Adaptiveness against Structure Adjustment}
\label{tab:structureequivalent}
\begin{tabular}{m{1.17cm}
<{\centering}|m{1.5cm}
<{\centering}|m{2.5cm}
<{\centering}|m{1.2cm}
<{\centering}}
\hline
\textbf{DNN}&\textbf{Variance}& \textbf{Accuracy of DNN-watermarked}&\textbf{HufuNet Accuracy} \\
\hline \hline
\multirow{3}{*}{\text{VGG11}}&\text{Original}&\text{77.77$\%$}&\text{92.32$\%$} \\
\text{}&\text{Reorder}&\text{77.77$\%$}& \text{18.15$\%$} \\
\text{}&\text{Restore}&\text{77.77$\%$}& \text{92.32$\%$} \\ \hline

\multirow{3}{*}{\text{Resnet18}}&\text{Original}&\text{86.14$\%$}& \text{92.32$\%$} \\
\text{}&\text{Reorder}&\text{86.14$\%$}& \text{13.71$\%$} \\
\text{}&\text{Restore}&\text{86.14$\%$}&  \text{92.32$\%$} \\  \hline

\multirow{3}{*}{\text{GoogLeNet}}&\text{Original}&\text{90.85$\%$}& \text{92.32$\%$} \\
\text{}&\text{Reorder}&\text{90.85$\%$}& \text{8.22$\%$} \\
\text{}&\text{Restore}&\text{90.85$\%$}& \text{92.32$\%$} \\ \hline

\end{tabular}
\end{table}

\begin{table}
\centering
\footnotesize
\caption{Adaptiveness against Parameter Adjustment}
\label{tab:parameterequivalent}
\begin{tabular}{m{1.17cm}
<{\centering}|m{1.5cm}
<{\centering}|m{2.5cm}
<{\centering}|m{1.2cm}
<{\centering}}
\hline
\textbf{DNN}&\textbf{Variance}& \textbf{Accuracy of DNN-watermarked}&\textbf{HufuNet Accuracy} \\
\hline \hline
\multirow{3}{*}{\text{VGG11}}&\text{Original}&\text{77.77$\%$}&\text{92.32$\%$} \\
\text{}&\text{PMCG}&\text{77.77$\%$}& \text{22.29$\%$} \\
\text{}&\text{Restore}&\text{77.77$\%$}& \text{92.32$\%$} \\ \hline
\multirow{3}{*}{\text{Resnet18}}&\text{Original}&\text{86.14$\%$}&\text{92.32$\%$} \\
\text{}&\text{PMCG}&\text{86.14$\%$}& \text{14.68$\%$} \\
\text{}&\text{Restore}&\text{86.14$\%$}& \text{92.32$\%$} \\ \hline
\multirow{3}{*}{\text{GoogLeNet}}&\text{Original}&\text{90.85$\%$}&\text{92.32$\%$} \\
\text{}&\text{PMCG}&\text{90.85$\%$}& \text{25.52$\%$} \\
\text{}&\text{Restore}&\text{90.85$\%$}& \text{92.32$\%$} \\ \hline
\end{tabular}
\begin{tablenotes}
\footnotesize
\item[1] PMCG refers to the DNN-watermarked whose parameters have been adjusted to launch functionality-equivalent attack. 
\end{tablenotes}
\end{table}


\subsection{Against Synthetic Attacks}
\label{subsec:combination adaptiveness with robustness}
After stealing a DNN model, the adversaries may try their best to undermine any watermarks embedded inside, e.g., utilizing fine-tuning, pruning, functionality-equivalent attack (structure/parameter adjustment and channel expansion). Below, we synthesized an advanced attack by combining the above attack methods to simulate the adversaries' best effort, and evaluate HufuNet against such a synthetic attack. First, we randomly choose two adjacent layers (4th, and 5th in our experiment) of VGG11 (with our watermark EPH embedded) to launch the channel expansion attack. Then, we fine-tune VGG11 as we did in Section \ref{finetunig} using the test dataset in 100 epochs. Afterward, we randomly select six layers to launch the functionality-equivalent attack. The structure adjustment is applied to two layers. The parameter adjustment is applied to another two layers, and both structure adjustment and parameter adjustment are applied to the other two layers. In the end, we prune it with the pruning rate of 10\%. For watermark detection, we retrieve $EPH_{retrieved}$ from the attacked model at its original embedding locations and merge $EPH_{retrieved}$ with our local SPH to obtain $Hufu_{combined}$. 

As shown in Table \ref{tab:advancedattack}, at the pruning rate of 10\%, the accuracy of $HufuNet_{combined}$ without restoring (``Direct'' in the table) drops to 10.23\%. After utilizing the restore approach in Section \ref{subsec:restore}, the accuracy of $Hufu_{combined}$ increases to 90.74\% with 100\% parameters correctly restored, thus allowing ownership verification. We also evaluate our restore approach against heavy pruning over the stolen model, up to 50\%. With the pruning rate of 30\%, HufuNet accuracy after restore is 83.95\%, still above the threshold for ownership verification (80\%), with 96.73\% of all embedded kernels correctly restored. The reasons that the restore rate is still above 96\% even at the pruning rate of 30\% are twofold. On the one hand, larger-value parameters contribute significantly to both the cosine similarity and SVD, while pruning starts to zeorize smaller-value parameters with larger-value ones unchanged. On the other hand, the cosine similarity and SVD compare the stolen model and the watermarked model at the granularity of channels, each of which includes a large number of parameters, i.e., $64*3*3$. Even at pruning rate of 30\%, there still exist lots of parameters (almost half on average) unchanged in most of the channels, resulting in higher cosine similarity and SVD values based on our experiments. At the pruning rate of 40\% however, the accuracy of HufuNet after restore is 51.25\% (below the threshold of 80\%), but the accuracy of the stolen VGG11 drops to 43.30\%, making it almost useless in performing its original task. Overall, based on the evaluation results, synthetic attacks launched by more capable adversaries can still be caught by our HufuNet watermark with the pruning rate up to 30\%. As the pruning rate keeps increasing, i.e., above 40\%, HufuNet fails ownership verification, but the adversaries end up with an almost useless stolen model (accuracy below 45\%).

\begin{table}
\begin{threeparttable}
\centering
\footnotesize
\caption{Against Synthetic Attacks}
\label{tab:advancedattack}
\begin{tabular}{m{1.5cm}
<{\centering}|m{1cm}
<{\centering}|m{1.1cm}
<{\centering}|m{1.1cm}
<{\centering}|m{1.9cm}
<{\centering}}
\hline

\multirow{2}{*}{\textbf{Pruning Rate}}&\multirow{2}{*}{\textbf{VGG11}}&\multicolumn{2}{c|}{\textbf{HufuNet Accuracy}}&\multirow{2}{*}{\textbf{Restored Rate}}\\
\cline{3-4}

\textbf{}& \textbf{}& \textbf{Direct}& \textbf{Restored}& \textbf{}\\
\hline \hline
\textbf{10\%}& \text{73.85\%}& \text{10.23\%}& \text{90.74\%}& \text{100.00\%}\\
\hline
\textbf{20\%}& \text{73.85\%}& \text{11.13\%}& \text{88.08\%}& \text{99.85\%}\\
\hline
\textbf{30\%}& \text{73.80\%}& \text{8.87\%}& \text{83.95\%}& \text{96.73\%}\\
\hline
\textbf{40\%}& \text{43.30\%}& \text{10.56\%}& \text{51.25\%}& \text{88.82\%}\\
\hline
\textbf{50\%}& \text{9.20\%}& \text{10.14\%}& \text{10.17\%}& \text{75.15\%}\\
\hline
\end{tabular}
\end{threeparttable}
\vspace{-7mm}
\end{table}


\subsection{Stealthiness}
\label{exp_stealthiness}
According to the results in Section \ref{subsec:Performance}, our watermark EPH only occupies a small portion of the DNNs-watermarked, thus difficult to be noticed. Although there exists a correlation (i.e., HMAC) between the embedding location in DNN-watermarked and the kernel parameters plus their indices in HufuNet, the strength of the key used to compute the HMAC and the secrecy of the indices in HufuNet make learning such correlation computationally hard. Hence, it is difficult for adversaries to learn the existence of the embedded watermark. Instead, they may resort to a statistical approach, e.g., examining the distribution of parameter values or their gradients of correct labels for any embedded watermark's footprint.

The adversaries with expertise in deep learning may have sufficient knowledge and understanding of their stolen model, e.g., roughly how the parameter values distribute. Therefore, we measure whether our watermark EPH will change the distribution of the parameter values in DNN-watermarked. We find that the distribution of the parameter distribution of DNN-to-be-protected (without EPH) and DNN-watermarked (with EPH) is too close to be distinguished, as shown in Figure \ref{fig:Distribution}(a) in Appendix. In addition, Grad-CAM~\cite{selvaraju2017grad} finds that the gradients of the score of true labels flowing back are global-average-pooled to obtain the neuron importance weights. The experienced adversaries are very likely to analyze the gradient distribution of neurons to discover any embedded watermark. Therefore, we choose 8,000 input images in the CIFAR-10 testing dataset, and for each correct label, compute the gradient of EPH and other parameters in DNN-watermarked. We find their distribution closely resembles each other, thus hard for adversaries to learn any embedded watermark evidence. The gradient distribution of EPH and other parameters in DNN-watermarked are compared in Figure \ref{fig:Distribution}~(b) in Appendix. 



\subsection{Security}
\label{subsec:security1}

An intuitive idea is that the adversary forges a valid HufuNet-style watermark from the public dataset $D_s$ and embeds it into DNN-watermarked. In particular, the adversary first trains a HufuNet from $D_s$; he/she then chooses a secret key and embeds EPH from the forged HufuNet into DNN-watermarked following the watermark embedding algorithm. In case of ownership dispute, the adversary may perform ownership verification on his/her stolen model, from which the forged watermark can be detected following the ownership verification algorithm. In such a case, the adversary's secret key would be different from the owner's secret key with an overwhelming probability. Therefore, it is highly unlikely that the adversary's forged watermark overwrites the owner's watermark in DNN-watermarked. Consequently, the adversary can only present a model with both his/her watermark and the owner's watermark detected, while the owner of the model can present the original DNN-watermarked with only the owner's watermark detected. It is thus clear who is the true owner to resolve the dispute.

Therefore, the adversary has to leverage existing parameters in the stolen model rather than revising any. Meanwhile, two requirements have to be satisfied to forge a valid and effective HufuNet-style watermark: (R1) the accuracy of the forged HufuNet passes the threshold; (R2) the positions where the embedded kernels (EPH) are retrieved should satisfy the correlation based on Line 3 in Algorithm \ref{alg:Position-injected}. Below we discuss and evaluate two different watermark forging attacks depending on which requirement to satisfy first: the first accuracy then correlation and first correlation then accuracy.


\noindent\textbf{First accuracy then correlation}
The adversary first trains his/her own HufuNet by designing a simple network structure and using the public training dataset. Once thoroughly tested, the adversary also divides it into two pieces and keeps the right piece as his/her SPH. Instead of embedding the left piece, the adversary searches the stolen model's parameter space for each of the convolution kernel of his/her EPH. An exact match of all the $3*3$ values of a kernel is hard to be found in general, so the adversary searches for closely-matched values, e.g., within the range of 25\% less and 25\% more of the searched value, which may still function similarly in the neural network. With those values found and retrieved (if lucky enough), the adversary can rebuild a $HufuNet_{combined}$ by combining them with the local SPH, and test its accuracy on the public training dataset. Note that the high accuracy of $HufuNet_{combined}$ is not enough. The adversary also needs to build the correlation between the embedding location and the parameter values of the kernel plus its original index in their HufuNet based on an appropriate key, according to Line 3 in Algorithm \ref{alg:Position-injected}.


We simulate the attack in two steps. In the first step, we train a HufuNet and check whether we can find a closely-matched kernel in DNN-watermarked (VGG11, Resnet18 and GoogLeNet) for each kernel in EPH, by setting the range as between below n\% and above n\% of the searched values\footnote{Based on our experience, a larger range makes it easy to find matched kernels, but cannot guarantee the high accuracy of $HufuNet_{combined}$, thus failing the ownership claim.}. However, even when n\% is set to be large, e.g., 200\%, the small values (e.g., 10e-7) in kernels are still difficult to be matched. Therefore, we ignore those small values when we perform kernel matches, since they generally contribute very little to the classification task of DNN models. The threshold to filter out the small values is chosen by setting the pruning rate to be 10\%, because the accuracy of HufuNet decreases significantly at a larger pruning rate. As shown in Table \ref{tab:securityofhufunet}, when we set the range n\% to be 25\%, lots of kernels in EPH cannot find matches in DNN-watermarked, i.e., total 64.90\%, 47.86\%, 52.92\% found in VGG11, GoogLeNet, and Resnet18, respectively. Until we increase the range to 75\%, 90\%, and 110\% for VGG11, GoogLeNet, and Resnet18 respectively, all the kernels of EPH can be matched. However, the accuracy of such $HufuNet_{combined}$ is quite low, i.e., 53.74\%, 60.96\%, and 34.00\% respectively, failing the adversary's goal of forging a valid watermark. In the second step, we show the computational difficulty for even a determined adversary to build the correlation for the closely-matched kernels by traversing all possible keys for HMAC function. It takes about $2.18*10^{-4}$ seconds to compute one HMAC operation (i.e., SHA256 in our experiment) using our server with Intel Xeon E5-2620 v4@2.10GHz CPU. According to~\cite{microsha256}, the recommended size of key for SHA256 is 64 bytes, which results in $2^{512}$ different keys for SHA256. Therefore, it takes $2.18*10^{-4}*2^{512}$ seconds for the adversary to traverse all the keys to try the correlation for one kernel in the forged HufuNet, which is computationally hard. Note that such brute force cannot ensure an appropriate key can always be found to build the correlation for all the kernels (i.e., 27,952) in the forged HufuNet. 

\noindent\textbf{First correlation then accuracy.}
Since both the training dataset $D_s$ and the testing dataset $D_{s\_TEST}$ are standardized for training and verifying HufuNet (see Section \ref{subsec: watermark_Genaration}), the adversaries cannot forge a valid HufuNet with arbitrary datasets. However, it does not rule out the possibility of forging a valid HufuNet from the public dataset $D_s$ and $D_{s\_TEST}$ with high accuracy. It is possible that the parameter space of DNN-watermarked is large and includes sufficient values required by an adversary to match a forged HufuNet. However, to prevent the adversaries from associating the matched values between DNN-watermarked and a forged HufuNet, our embedding algorithm correlates each EPH kernel's embedding location in DNN-watermarked with a secret key, the parameter values of the EPH kernel, and its index through a secure HMAC function.  

Since Algorithm \ref{alg:Position-injected} is public, the adversary may randomly select a key, test each kernel in the stolen model based on Line 3 in Algorithm \ref{alg:Position-injected}, and try to find all the kernels that satisfy the correlation. The adversary can explore quite several different keys, e.g., 1,000, and check which key finds the most kernels that satisfy the correlation. If enough such kernels could be found based on a particular key, the adversary can claim those kernels are his/her EPH. To forge a valid HufuNet-style watermark, the adversary simply places those kernels into the convolution layers, randomly initializes a fully connected layer, and trains the fully connected layer using the public dataset $D_s$. Such watermark might have high accuracy on the public testing dataset as well. Although theoretically plausible, our experiment below shows that it is computationally tricky for the adversary to discover an appropriate secret key to find him/her enough kernels satisfying the correlation. 

We train VGG11 with our HufuNet watermark embedded, as DNN-watermarked. After stealing the model, the adversaries can try a random key on all the 1,024,192 kernels in VGG11, to find those kernels that satisfy the correlation as Line 3 in Algorithm \ref{alg:Position-injected}. We simulate this attack by testing 100 random keys, and find at most 3 kernels that can satisfy the desired correlation. With such a limited number of kernels, it is impossible to build convolutional layers of any DNN model. Brute force searching of all keys is computationally hard based on our evaluation. In the worst case, the adversary needs $2.18*10^{-4}*2^{512}$ seconds to traverse all the keys for just one kernel as presented above. Note that brute force searching of all possible keys may let the adversary obtain our key used to embed our HufuNet, but it is computationally hard to traverse all the keys (a key of 64 byte long means 2$^{256}$ different keys in total.).

\begin{table}
\begin{threeparttable}
\centering
\footnotesize
\caption{Security of $HufuNet$}
\label{tab:securityofhufunet}
\begin{tabular}{m{1.15cm}
<{\centering}|m{0.75cm}
<{\centering}|m{0.75cm}
<{\centering}|m{0.75cm}
<{\centering}|m{0.75cm}
<{\centering}|m{0.6cm}
<{\centering}|m{1cm}
<{\centering}}
\hline
\multirow{2}{*}{\textbf{Models}}&\multicolumn{5}{c|}{\textbf{Found Percentage at Different Thresholds}}&\multirow{2}{*}{\parbox{3cm}{\textbf{ HufuNet \\ Accuracy$^{1}$}}}\\
\cline{2-6}
\text{ }& \text{25\%}&\text{50\%}& \text{75\%}& \text{80\%}& \text{90\%} \\
\cline{2-7} \hline \hline
\textbf{VGG11}& \text{64.90$\%$}& \text{97.80$\%$} &\text{100$\%$}& \text{100$\%$}& \text{100$\%$}& \text{53.74$\%$} \\ \hline
\textbf{Resnet18}& \text{52.93$\%$}&  \text{93.24$\%$} &\text{99.99$\%$}& \text{100$\%$}& \text{100$\%$}& \text{34.00$\%$} \\ \hline
\textbf{GoogLeNet}& \text{47.86$\%$}& \text{87.76$\%$} &\text{99.84$\%$}& \text{99.97$\%$}& \text{100$\%$}& \text{60.96$\%$} \\ \hline
\end{tabular}
\begin{tablenotes}
\item[1] HufuNet Accuracy is measured when 100\% kernels are found at the lowest range for each model.
\end{tablenotes}
\end{threeparttable}
\end{table}

\vspace{-10pt}
\subsection{Integrity}
\label{subsec:integrity}
We evaluate the integrity of HufuNet by testing its ownership verification on innocent models not copied or retrained from ours. In this experiment, we intend to use the innocent models that always share the same network structure and the same task as ours, thus highly resembling our watermarked models. In particular, we consider two different kinds of such innocent models: (Innocent 1) the same training dataset as DNN-watermarked and (Innocent 2) a different training dataset than DNN-watermarked. We train each innocent model for 100 epochs, and each epoch begins with different initialization parameters to ensure high accuracy. We use the models obtained above as the innocent models, without embedding our watermark inside them. Upon ownership verification, we build $HufuNet_{combined}$ by combining the local SPH and the $EPH_{retrieved}$ from the innocent models based on the computed embedding locations. The experimental results are shown in Table \ref{tab:integrity}. The accuracy of $HufuNet_{combined}$ is 11.98\% for Innocent 1 and 9.97\% for Innocent 2 on average. Therefore, none of the innocent models can be falsely claimed as ours. For other models with different structures, main tasks, or training datasets, it is unlikely for HufuNet to claim ownership falsely.

\begin{table}
\begin{threeparttable}
\centering
\footnotesize
\caption{Integrity of HufuNet}
\label{tab:integrity}
\begin{tabular}{m{1.5cm}
<{\centering}|m{0.75cm}
<{\centering}|m{1.25cm}
<{\centering}|m{0.95cm}
<{\centering}|m{0.95cm}
<{\centering}|m{0.75cm}
<{\centering}}
\hline
\textbf{ }& \textbf{VGG11}& \textbf{GoogLeNet}& \textbf{Resnet18}& \textbf{Resnet34}& \textbf{LSTM}\\
\hline \hline
\textbf{Innocent 1$^{1}$}& \text{13.06$\%$} &\text{14.39$\%$}& \text{10.02$\%$}& \text{12.21$\%$}& \text{10.21$\%$} \\ \hline
\textbf{Innocent 2$^{2}$}& \text{10.59$\%$} &\text{10.00$\%$}& \text{9.27$\%$}& \text{10.00$\%$}& \text{9.99$\%$} \\ \hline
\end{tabular}
\begin{tablenotes}
\item[1] Innocent 1 refers to the innocent model with the same architecture and the same task trained with the same dataset as ours.
\item[2] Innocent 2 refers to the innocent model with the same architecture and the same task but trained with different dataset than ours.
\end{tablenotes}
\end{threeparttable}
\end{table}

\subsection{Fidelity}
\label{subsec:fidelity}
Ideally, our watermark EPH should have a negligible impact on the performance of the DNN-to-be-protected. 
To demonstrate such fidelity, we perform experiments on the models trained on datasets as in Table \ref{tab:parameter} and evaluate the accuracy of DNNs-to-be-protected and DNNs-watermarked. Table~\ref{tab:fidelity} shows that the accuracy of DNNs-watermarked is always quite close to that of DNNs-to-be-protected. We believe it is mainly because our watermark EPH only occupies a small portion of the entire parameters of DNNs-watermarked and participates in the training of DNNs-watermarked, thus incurring little impact on their accuracy. 


\begin{table}
\begin{threeparttable}
\centering
\footnotesize
\caption{Fidelity of HufuNet}
\label{tab:fidelity}
\begin{tabular}{m{1.4cm}
<{\centering}|m{0.8cm}
<{\centering}|m{1.25cm}
<{\centering}|m{0.95cm}
<{\centering}|m{0.95cm}
<{\centering}|m{0.75cm}
<{\centering}}
\hline
\textbf{ }& \textbf{VGG11}& \textbf{GoogLeNet}& \textbf{Resnet18}& \textbf{Resnet34}& \textbf{LSTM}\\
\hline \hline
\textbf{w/o EPH$^{1}$} & \text{77.16$\%$}& \text{90.71$\%$}& \text{85.67$\%$}& \text{69.98$\%$}& \text{87.35$\%$}\\ \hline
\textbf{w/ EPH$^{2}$}& \text{77.77$\%$} &\text{90.85$\%$}& \text{86.14$\%$}& \text{70.20$\%$}& \text{87.08$\%$} \\ \hline
\textbf{Performance change$^{3}$}& \text{+0.61$\%$} &\text{+0.14$\%$}& \text{+0.47$\%$}& \text{+0.22$\%$}& \text{-0.27$\%$} \\ \hline
\end{tabular}
\begin{tablenotes}
\item[1] w/o EPH represents the DNN-to-be-protected, without our EPH. 
\item[2] w/ EPH indicates DNN-watermarked, with our EPH.
\item[3] The performance change of the watermarked model compared with the original model. The positive value indicates performance improvement.
\end{tablenotes}
\end{threeparttable}
\end{table}

\subsection{Comparison with State-of-the-art}
\label{subsec:comparison}
We compare our HufuNet with the state-of-the-art watermark approach~\cite{adi2018turning} mainly in the aspect of model fine-tuning. Firstly, we train Resnet18 with the same hyperparameters and the same dataset using the same pre-processing (i.e., normalization, random cropping) as in ~\cite{adi2018turning}. Then, regarding fine-tuning the model, according to~\cite{chen2019leveraging}, setting a larger initial learning rate and then properly decreasing it will effectively destruct the embedded watermark without compromising the performance of the model. So we use a larger initial learning rate of 0.05 and decrease it to 1e-7 gradually after 100 epochs by cosine annealing. To fine-tune the DNNs-watermarked using the backdoor approach \cite{adi2018turning} and our HufuNet, respectively. The two watermarks embedded into the Resnet18 model are fine-tuned using 8,000 images and verified using another 2,000 images from the CIFAR-10 testing dataset. Table \ref{tab:Contrastblackfinetuning} shows the experimental results. We find that the fine-tuning does not impact the accuracy of the watermarked models for both approaches. However, the accuracy of their backdoor based watermark~\cite{adi2018turning} drops significantly from the beginning of fine-tuning and continues decreasing as the number of epochs increases. In contrast, the accuracy of HufuNet tends to be stable, always above 92\%, resulting in 100\% ownership verification (as in Section \ref{subsec:Performance}, with $\tau_{acc} = 15\%$, above 80\% HufuNet accuracy indicates ownership verification). We also evaluate the two watermark approaches against pruning. With the pruning rate up to 60\%, we find both approaches demonstrate good robustness against pruning, with 100\% ownership verification.

\begin{table}
\begin{threeparttable}
\centering
\footnotesize
\caption{Comparison of Fine-tuning Performance}
\label{tab:Contrastblackfinetuning}
\begin{tabular}{m{0.7cm}
<{\centering}|m{1.24cm}
<{\centering}|m{1.1cm}
<{\centering}|m{1.24cm}
<{\centering}|m{1.0cm}
<{\centering}|m{1cm}
<{\centering}}
\hline
\multirow{2}{*}{\textbf{\makecell[c]{Epochs}}} & \multicolumn{2}{c|}{\textbf{Backdoored WM \cite{adi2018turning}}}& \multicolumn{3}{c}{\textbf{HufuNet}}\\ \cline{2-6}

\text{}&\textbf{Main task}&\textbf{\makecell[c]{SR}}&\textbf{Main task}& \textbf{Accuracy} & \textbf{SR} \\
\hline \hline
\textbf{10} & \text{91.00$\%$} &\text{42.00$\%$}& \text{92.85$\%$}&\text{92.32$\%$}&\text{100$\%$}\\ \hline
\textbf{20} & \text{91.15$\%$} &\text{37.00$\%$}& \text{92.65$\%$}&\text{92.34$\%$} &\text{100$\%$}\\ \hline
\textbf{30}& \text{91.05$\%$} &\text{35.00$\%$}& \text{92.65$\%$}&\text{92.38$\%$} &\text{100$\%$}\\ \hline
\textbf{40}& \text{90.95$\%$} &\text{29.00$\%$}& \text{92.45$\%$}&\text{92.36$\%$} &\text{100$\%$}\\ \hline
\textbf{50}& \text{90.85$\%$} &\text{25.00$\%$}& \text{92.40$\%$}&\text{92.38$\%$} &\text{100$\%$}\\ \hline
\textbf{60}& \text{90.65$\%$} &\text{26.00$\%$}& \text{92.35$\%$}&\text{92.36$\%$} &\text{100$\%$}\\ \hline
\textbf{70}& \text{90.65$\%$} &\text{26.00$\%$}& \text{92.35$\%$}&\text{92.38$\%$} &\text{100$\%$}\\ \hline
\textbf{80}& \text{90.65$\%$} &\text{26.00$\%$}& \text{92.45$\%$}&\text{92.37$\%$} &\text{100$\%$}\\ \hline
\textbf{90}& \text{90.85$\%$} &\text{25.00$\%$}& \text{92.45$\%$}&\text{92.37$\%$} &\text{100$\%$}\\ \hline
\textbf{100}& \text{90.85$\%$} &\text{25.00$\%$}& \text{92.45$\%$}&\text{92.37$\%$} &\text{100$\%$}\\ \hline
\end{tabular}
\begin{tablenotes}
\item SR indicates success rate that the watermark can verify ownership correctly.
\end{tablenotes}

\end{threeparttable}
\vspace{-4mm}
\end{table}

\vspace{-4pt}
\section{Related Work}
\label{sec:Related work}
\vspace{2pt}\noindent\textbf{Watermarking neural networks.} 
Uchida et al.~\cite{uchida2017embedding} proposed the idea of utilizing a parameter regularizer to embed a watermark in the parameter space of the model and retrieved it during ownership verification. However, such an approach is vulnerable to statistical analysisattack~\cite{wang2019attacks},  because the layer's parameter variance with embedded watermark is more significant than other layers without watermark. In contrast, our EPH was embedded into nearly every layer of DNN-watermarked and occupied only a small portion of its parameter space, so the variances between the layers with and without EPH are not entirely different, thus ensuring stealthiness. 


Recently, backdoor-based watermarking approaches have been proposed to probe black-box or grey-box neural network models that are suspects of IP infringement~\cite{adi2018turning}, \cite{zhang2018protecting}, \cite{guo2018watermarking}, \cite{rouhani2018deepsigns}. Adi et al.~\cite{adi2018turning} embedded a set of abstract images with predefined labels into DNN models to build the trigger set, which acted as digital watermarks (only known to model owners) against the suspect models. Zhang et al.~\cite{zhang2018protecting} enriched the training dataset by adding content, noise, or irrelevant class examples as triggers, and leveraged them as model watermarks. Guo et al.~\cite{guo2018watermarking} proposed a watermarking approach similar to~\cite{zhang2018protecting} that can be used on embedded devices.
However, the backdoor watermark approaches proposed above are non-blind watermark, which either fails to defend against evasion attacks~\cite{hitaj2018have} or does not explicitly address the problem of fraudulent ownership claim. Evasion attack~\cite{hitaj2018have} evades backdoor-based watermark verification by training a powerful detector to identify key samples in the trigger set of owners, and returning a random label rather than the predefined label when it detects a key sample used as input. 

To overcome the ``non-blind'' problem, backdoor-based blind watermarking approach \cite{namba2019robust}, \cite{li2019prove} has been proposed. Namba et al.~\cite{namba2019robust} utilized the original samples in the training dataset as watermarks by changing their labels. Li et al.~\cite{li2019prove} built watermarks by integrating a specific logo into the original samples and changing their labels. Merrer et al.~\cite{le2019adversarial} proposed a zero-bit watermarking algorithm based on adversarial examples. It fine-tunes the DNN-to-be-protected, so that the produced DNN model contains certain true/false adversarial examples, which can be used as secret keys to achieving ownership verification of the DNN models. These approaches still cannot deal with the fraudulent claims of ownership, since attackers can still pretend to be owners by utilizing adversarial samples to build their own trigger set. Different from these studies, our HufuNet can achieve high robustness and security.


\vspace{2pt}\noindent\textbf{Counterattacks against watermarking.} Uchida et al. \cite{uchida2017embedding} proposed to fine-tune or prune the DNN model to modify its parameters, thus removing the embedded watermark.
Wang et al.~\cite{wang2019attacks} proposed to reveal any watermark based on the change of statistical distribution of parameters of the layer where the watermark is embedded. Yang et al.~\cite{yang2019effectiveness} demonstrated that distillation-based attacks can be effective to perform watermark elimination, based on the premise that the adversaries have plenty of data to fine-tune the model. Chen et al.~\cite{chen2019leveraging} proved that the well-designed fine-tuning method can remove the watermark without compromising the function of the model with a large learning rate and unlabeled data. Hitaj et al.~\cite{hitaj2018have} proposed a method of evasion attacks against backdoor watermarks, which evade the homeowner's verification, thereby avoiding the discovery of model theft.
Although these counterattacks significantly threaten the backdoor-based watermarking, our design has better robustness against them.

\vspace{-14pt}
\vspace{12pt}
\section{Conclusion}
\label{sec:Conclusion}

In this paper, we proposed HufuNet, a robust and secure watermarking approach, to protect DNN models' ownership. We divided HufuNet into two pieces, with the left piece EPH embedded into the DNN-to-be-protected and the right piece SPH kept locally as the secret to verify ownership of suspect DNN models. Compared with previous watermark approaches, the ownership verification of HufuNet is more robust against model fine-tuning, pruning and is more secure against watermark forging, which is demonstrated by our experimental results conducted on five popular DNN models. Meanwhile, our watermarking approach has a negligible impact on the performance of the DNN models it protects. It is robust to kernel cutoff/supplement, adaptive to the functionality-equivalent attack, and does not falsely claim ownership on innocent models.

\bibliographystyle{ACM-Reference-Format}
\bibliography{reference.bib}


\renewcommand{\thesubsection}{\Alph{subsection}}
\clearpage

\section*{Appendix}
\label{sec:Appendix}

\subsection{Analysis of Cosine Similarity.} \label{analysis_of_channel_expansion_principle}
Cosine similarity is to evaluate the similarity of two vectors by calculating the cosine value of the angle between them. The cosine similarity for the two vectors $A$ and $B$ is calculated as follows:

\begin{equation}
similarity = \frac{A \cdot B}{ \|A\|  \|B \|}
\end{equation}
We use cosine similarity to restore the order of output channels of each layer, supposing the structure adjustment has been applied on the model. If the parameters on each output channel do not change, the restore works right away by locating those identical channels (with cosine similarity as one). However, advanced attacks may involve parameter adjustment of the model as well, e.g., fine-tuning, hoping to undermine the embedded watermarks effectively. In such a scenario, we can still restore correctly, since the cosine similarity of the same output channel before fine-tuning and after fine-tuning is significantly larger than that of different output channels based on our observation as below. Once the output channel of one layer is restored correctly, the input channel of the next layer can be restored correspondingly due to their connections. Below, we focus on discussing the cosine similarity of the output channels.

\begin{figure}[htbp]
\centering

\subfigure[VGG]{
\begin{minipage}[t]{0.5\linewidth}
\centering
\includegraphics[width=1.65in]{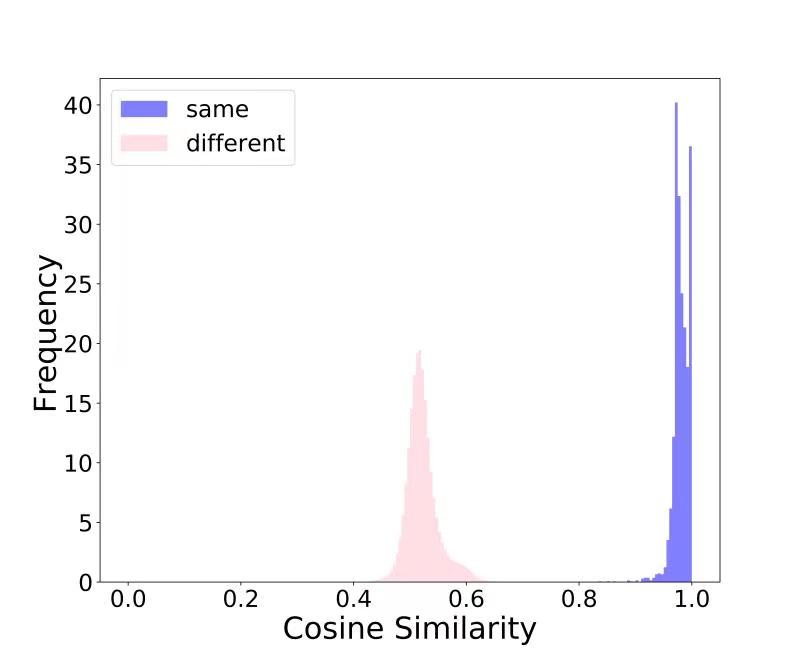}
\end{minipage}%
}%
\subfigure[GoogLeNet]{
\begin{minipage}[t]{0.5\linewidth}
\centering
\includegraphics[width=1.65in]{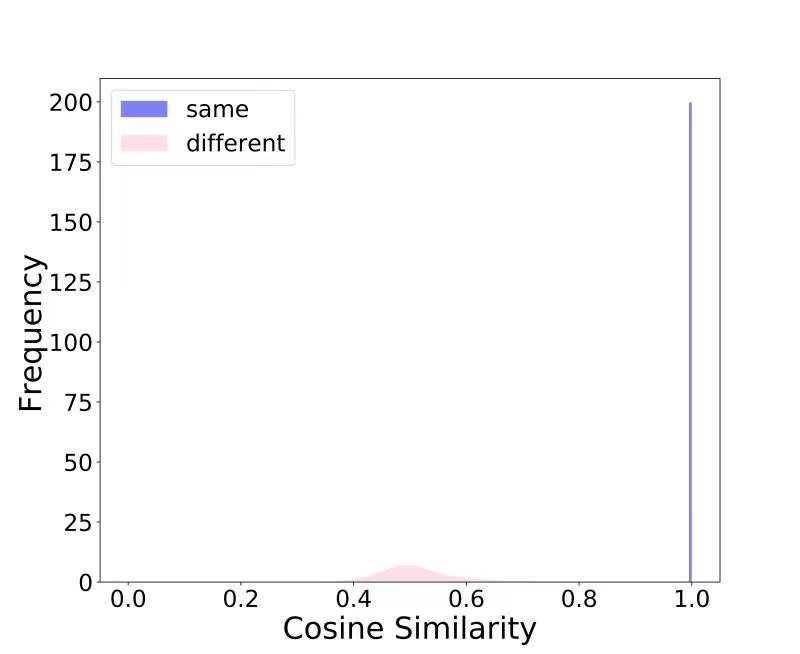}
\end{minipage}%
}%
\quad

\subfigure[Resnet18]{
\begin{minipage}[t]{0.5\linewidth}
\centering
\includegraphics[width=1.65in]{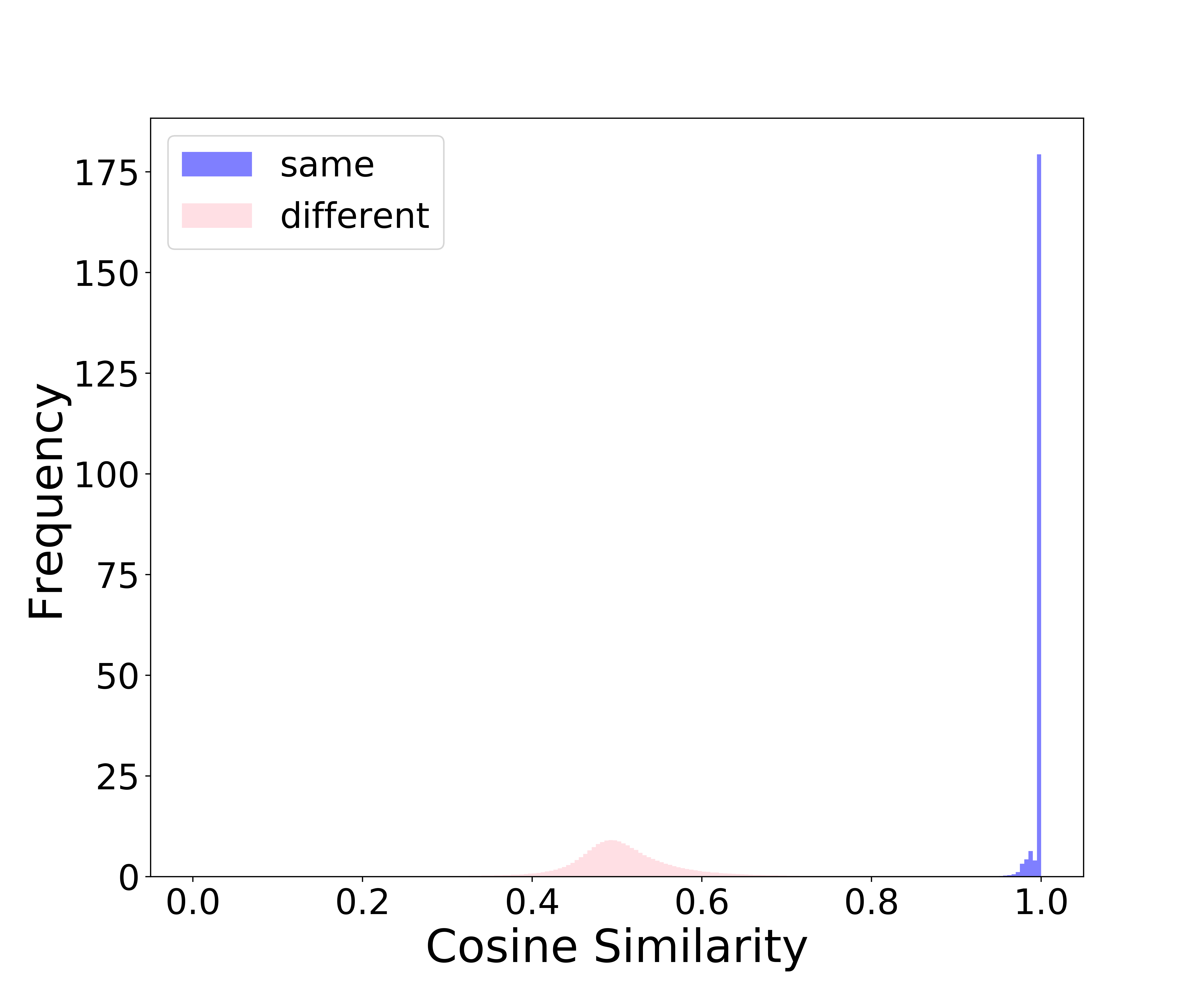}
\end{minipage}
}%
\subfigure[Resnet34]{
\begin{minipage}[t]{0.5\linewidth}
\centering
\includegraphics[width=1.65in]{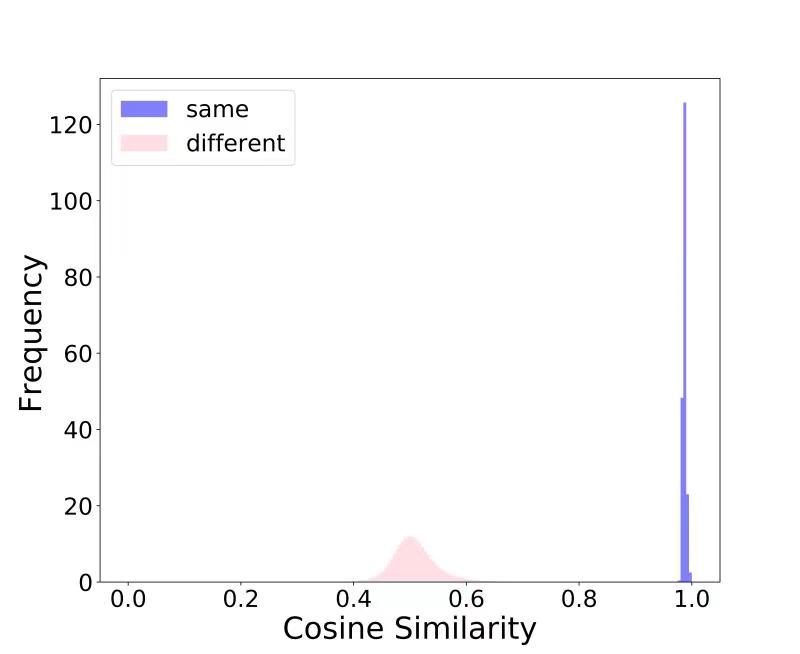}
\end{minipage}
}%

\centering
\caption{Distribution of Cosine Similarity of the Same Output Channel vs Different Output Channels}
\label{analysis_cosine_evaluation}
\end{figure}

Given a neural network model $A$, we can obtain $A'$ after fine-tuning $A$. Therefore, an output channel $OC$ in $A$ becomes $OC'$ in $A'$. To retrieve $OC'$ from $A'$ correctly based on $OC$, the cosine similarity of $OC$ and $OC'$ (referred to as the same output channel) should be distinguishable from those of $OC$ and all other output channels (except $OC'$) in $A'$ (referred to as different output channels). We demonstrate the distinction via the following experiments. We first choose four models, i.e., VGG11, GoogLeNet, Resnet18 and Resnet34. These four models are used in the fine-tuning experiment in Section \ref{subsec:Robustness}. 
Then, we fine-tune all the models with the same setting as Section \ref{subsec:Robustness}. Hence, the cosine similarity of the same output channel and different output channels can be computed for each model as shown in Figure \ref{analysis_cosine_evaluation}. The evaluation results show that for all the models, the cosine similarity of the same output channel is significantly larger compared with those of different output channels. The former is with the mean normal distribution N (0.9766, 0.0097), which means 99.74\% of values are distributed in (0.9473, 1), while the latter is with the mean normal distribution N (0.5123,0.0526), which means 99.74\% of values are distributed in (0.3546, 0.6670). Therefore, we believe cosine similarity approach can effectively restore the order of the output channels and retrieve the EPH correctly even after fine-tuning.

\vspace{-8pt}
\subsection{Analysis of Channel Expansion.} 
The adversaries can randomly choose two adjacent layers from their stolen model, indicated by the matrices $A$ and $B$. Then they increase the output dimension of $A$ (by increasing its output channels by $A_{inc}$) to get $A'$, and $B$ (by increasing its input channels by $B_{inc}$) to get $B'$. To ensure functionality-equivalent, $A_{inc}$ and $B_{inc}$ should be carefully chosen, so that $(A' * B') x$ = $(A * B) x$, where $x$ is the feature map propagated in neural networks. It implies that $(A_{inc} * B_{inc}) = 0$, which means that $A_{inc}$ and $B_{inc}$ have no contribution to the calculation of feature map $x$. 

Let $A$ be a $m * n$ matrix, $B$ be a $n * l$ matrix; $A_{inc}$ be a $m * k$ matrix; and $B_{inc}$ be a $k * l$ matrix. From the equation $(A_{inc} * B_{inc}) = 0$, we have $mk+kl$ variables and $ml$ linear equations. If $ml \ge mk+kl$, we may have a unique solution. Since zero for each variable is always a solution, the unique solution should be zeros for all the $mk+kl$ variables. Such distinguishable $A_{inc}$ and $B_{inc}$ are easy to be filtered out during restore. If $ml < mk+kl$, the solution is not unique. In the worst case, attacker can choose a big $k$, thus having multiple solutions, which are functions of $mk+kl-ml$ free variables. This means that attacker can choose $mk+kl-ml$ variables freely, while other $ml$ variables are determined by these $mk+kl-ml$ variables. 

Certainly the adversaries would prefer the scenario of many solutions for $A_{inc}$ and $B_{inc}$ (i.e., $mk+kl-ml$ free variables), hoping to find one that can fail our restore approach. With a random solution, the channels of $A_{inc}$ and $B_{inc}$ cannot result in higher cosine similarity values with those in the EPH as demonstrated in Appendix \ref{analysis_cosine_evaluation}. Without knowing our embedded EPH, the adversaries cannot intend to find a solution containing channels with higher cosine similarity values with those in the EPH. 
In Section \ref{evaluation_channel_expansion} and Section \ref{subsec:combination adaptiveness with robustness}, we evaluate channel expansion attack by generating random parameters for $A_{inc}$ and $B_{inc}$ with the restriction of $(A_{inc} * B_{inc}) = 0$ to meet the functionality-equivalent requirement.

\begin{figure*}[ht]
\centering
\epsfig{figure=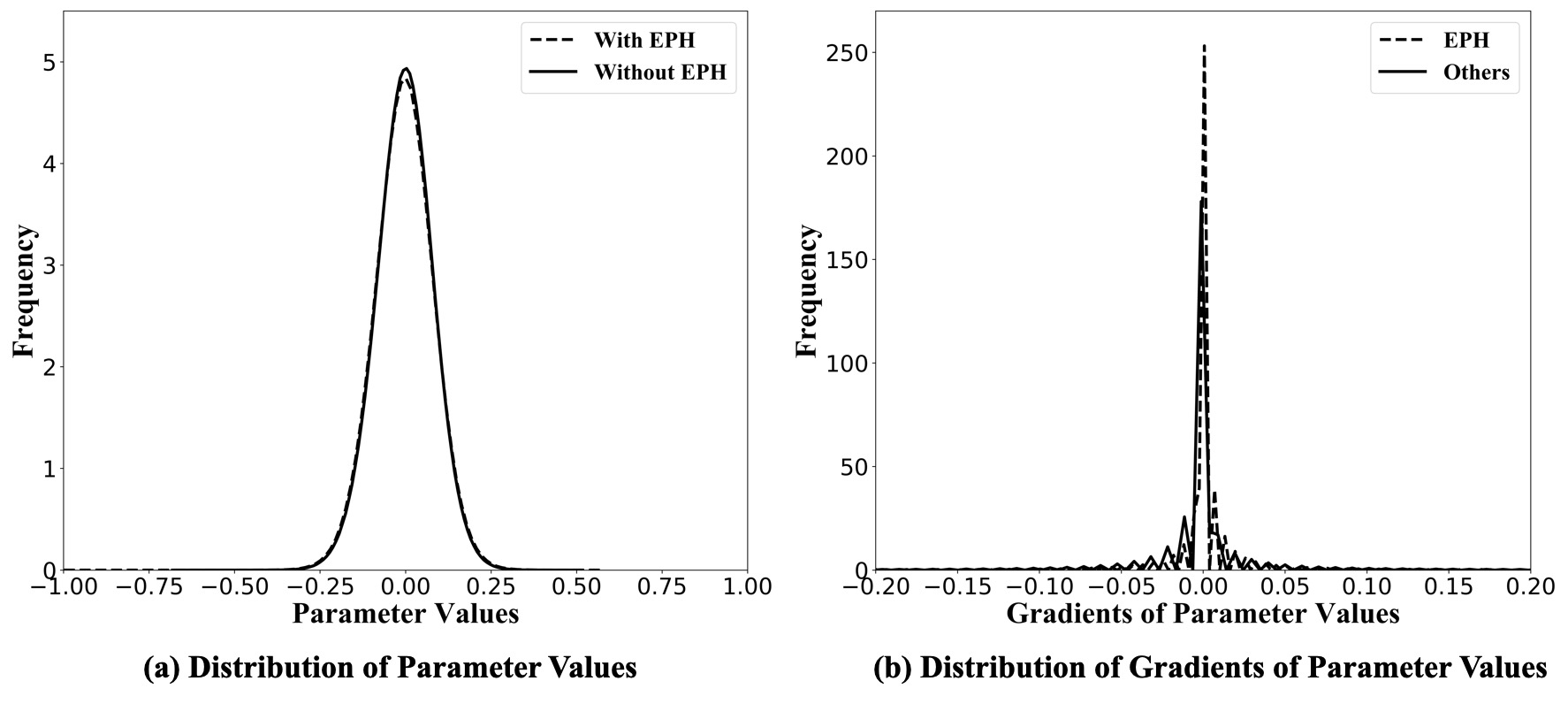, width=0.85\linewidth} 
\caption{Distribution of Parameter Values and Distribution of Gradients of Parameter Values}
\label{fig:Distribution}
\end{figure*}

\begin{table*}[h]

\centering
\begin{threeparttable}
\footnotesize
\caption{Impact of Embedding Backdoors as Watermark}
\label{tab:numberofbackdoor}
\begin{tabular}{m{4cm}
<{\centering}|m{4cm}
<{\centering}|m{4cm}
<{\centering}}
\hline
\textbf{Number of Backdoor Samples$^{1}$}& \textbf{Accuracy of Watermarked VGG11} & \textbf{Accuracy of Watermarked Resnet18~\cite{adi2018turning}}\\
\hline \hline
\text{$0$} & \text{77.32$\%$} &\text{93.42$\%$} \\ \hline
\text{$100$}& \text{76.49$\%$} &\text{93.65$\%$} \\ \hline
\text{$200$}& \text{75.48$\%$} &\text{92.11$\%$} \\ \hline
\text{$300$}& \text{75.48$\%$} &\text{92.14$\%$} \\ \hline
\text{$400$}& \text{75.92$\%$} &\text{91.74$\%$} \\ \hline
\end{tabular}
\begin{tablenotes}
\item[1] Number of backdoor samples indicates the number of samples in the trigger set that activates a specific backdoor embedded into the model as watermark. 
\end{tablenotes}
\end{threeparttable}
\end{table*}

\begin{table*}[h]
\centering
\begin{threeparttable}

\footnotesize
\caption{Datasets for DNN-watermarked and HufuNet}
\label{tab:datasets}
\begin{tabular}{m{3.0cm}
<{\centering}|m{5.0cm}
<{\centering}|m{4.0cm}
<{\centering}}
\hline
\textbf{Models} & \textbf{Datasets for DNN-watermarked$^{1}$}& \textbf{Dataset for HufuNet}\\
\hline \hline
\textbf{VGG11} & \text{CIFAR-10} &\text{FASHION-MNIST} \\ \hline
\textbf{GoogLeNet}& \text{CIFAR-10} &\text{FASHION-MNIST}\\ \hline
\textbf{Resnet18}& \text{CIFAR-10} &\text{FASHION-MNIST} \\ \hline
\textbf{Resnet34}& \text{CIFAR-100} &\text{FASHION-MNIST} \\ \hline
\textbf{LSTM}& \text{IMDB} &\text{FASHION-MNIST} \\ \hline
\end{tabular}
\begin{tablenotes}

\item[1] DNN-watermarked (with EPH embedded into DNN-to-be-protected) is the actual model we need to train on the dataset.

\end{tablenotes}
\end{threeparttable}
\end{table*}

\begin{table*}[h]
\centering
\begin{threeparttable}
\footnotesize
\caption{Parameter Sizes}
\label{tab:parameter}
\begin{tabular}{m{2.8cm}
<{\centering}|m{1.4cm}
<{\centering}|m{2.0cm}
<{\centering}|m{1.5cm}
<{\centering}|m{1.5cm}
<{\centering}|m{1.4cm}
<{\centering}}
\hline
\textbf{ }& \textbf{VGG11}& \textbf{GoogLeNet}& \textbf{Resnet18}& \textbf{Resnet34}& \textbf{LSTM}\\
\hline \hline
\textbf{Before embedding} & \text{36028KB} &\text{23056KB}& \text{43712KB}& \text{83425KB}& \text{9251KB} \\ \hline
\textbf{HufuNet}& \multicolumn{5}{c}{1241KB} \\ \hline
\textbf{EPH}& \multicolumn{5}{c}{1238KB}\\ \hline
\textbf{Embedding percentage}& \text{3.44\% } &\text{5.37\% }& \text{ 2.83\%}& \text{1.48\% }&\text{13.38\%}\\ \hline
\end{tabular}

\end{threeparttable}
\end{table*}

\end{document}